\documentclass[12pt]{article}
\pdfoutput=1
\usepackage{xcolor}
\usepackage{hyperref}
\usepackage{subcaption}
\usepackage[utf8]{inputenc}

\usepackage{setspace}
\usepackage{amsmath, amssymb, amsthm, float, graphicx}
\numberwithin{equation}{section}

\hypersetup{colorlinks=false, linkcolor=blue, citecolor=red}

\def\beq{\begin{eqnarray}}\def\eeq{\end{eqnarray}}
\def\be{\begin{equation}}\def\ee{\end{equation}}
\def\g{\gamma}

\def\a{\alpha}
\def\e{\epsilon}
\def\k{\kappa}
\def\b{\beta}
\def\d{\delta}
\def\c{\chi}

\def\D{\Delta}
\def\G{\Gamma}

\def\pd{\partial}
\def\tq{\tilde{q}}
\def\ta{\tau}
\def\bz{\bar{z}}

\def\G{\Gamma}
\def\mc{{\mathcal{C}}}
\def\bh{\bar{h}}

\begin{document}
\title{\bf Analytic bootstrap at large spin }
\date{}

\author{Apratim Kaviraj\footnote{apratim@cts.iisc.ernet.in}, ~Kallol Sen\footnote{kallol@cts.iisc.ernet.in} ~and Aninda Sinha\footnote{asinha@cts.iisc.ernet.in}\\ ~~~~\\
\it Centre for High Energy Physics,
\it Indian Institute of Science,\\ \it C.V. Raman Avenue, Bangalore 560012, India. \\}
\maketitle

\abstract{We use analytic conformal bootstrap methods to determine the anomalous
dimensions and OPE coefficients for large spin operators in general
conformal field theories in four dimensions containing a scalar operator
of conformal dimension $\Delta_\phi$. It is known that such theories will contain an
infinite sequence of large spin operators with twists approaching
$2\Delta_\phi+2n$ for each integer $n$. By considering the case where such
operators are separated by a twist gap from other operators at large spin,
we analytically determine the $n$, $\Delta_\phi$ dependence of the anomalous dimensions. We find that for all $n$, the anomalous dimensions are negative for $\Delta_\phi$ satisfying the unitarity bound. We further compute the first subleading correction at large spin and show that it becomes universal for large twist.  In the limit when $n$ is large, we find exact agreement with the AdS/CFT prediction corresponding to the Eikonal limit of a 2-2 scattering with dominant graviton exchange.}

\vskip 1cm

\tableofcontents

\onehalfspacing

\section{Introduction}

Over the last few years there has been a resurgent interest in conformal bootstrap methods \cite{Rattazzi},\cite{Polchinski},\cite{Hogervorst} using the seminal work on conformal blocks by Dolan and Osborn \cite{Dolan}. Using numerical methods, interesting constraints have been placed on conformal field theories in diverse dimensions \cite{Rattazzi}. Applications have been found in diverse field theories ranging from supersymmetric conformal field theories \cite{SCFTs} to the 3d-Ising model at criticality \cite{Showk}. The lessons learnt using these methods transcend any underlying Lagrangian formulation and are hoped to be very general. Our aim in this paper is to present new analytic results for conformal field theories in four dimensions.

Analytic bootstrap methods have been used in \cite{anboot,Komargodski} to study the four point function of four identical scalar operators. It has been shown that there must exist towers of operators at large spins with twists $2\D_\phi+2n$ with $\D_\phi$ being the conformal dimension of the scalar and $n\geq 0$ is an integer.  For the case where a single tower of operator exists with twists $2\D_\phi+2n$  and there is a twist gap between these operators and any other operator, one can calculate the anomalous dimensions of such operators. In four dimensions, the anomalous dimensions in the large spin $(\ell\gg1)$ limit for these operators for $n=0$ are given by \cite{anboot, Komargodski, Alday2},
\be
\g(0,\ell)=-\frac{c_0}{\ell^2}\,,
\ee
where $c_0>0$. This conclusion is consistent with the Nachtmann theorem \cite{Nachtmann:1973mr} which predicts that the leading operators at a given $\ell$ should have twists increasing with $\ell$. However it is not known if this behaviour persists for arbitrary $n$ introduced above (for a recent study\footnote{In \cite{Vos} the dependence of $n$ in the limit $\ell \gg n \gg 1$ is extracted numerically from a recursion relation but from that approach it is not possible to make general conclusions. After our paper appeared on the arXiv, \cite{Vos} furthered the analysis to agree perfectly with our findings.} see \cite{Vos}). 

Recently it has been pointed out that in the context of the AdS/CFT correspondence, there is a connection between the CFT anomalous dimensions and the bulk Shapiro time delay \cite{Cornalba1, Cornalba2,Cornalba3,Camanho:2014apa}. In \cite{Camanho:2014apa} it was argued that to preserve causality, the Shapiro time delay should be positive and hence the anomalous dimensions of double trace operators negative. Thus it is of interest to see what happens to $\g(n,\ell)$ for $n>0$. In the literature, it has been shown using input from AdS/CFT that using the results for the four point functions of dimension-2 and dimension-3 half-BPS multiplets in $\mathcal{N}=4$ supersymmetric SU($N$) Yang-Mills theories, to leading order in $1/N^2$,  $\gamma(n,\ell)\leq 0$ for all $n$--see \cite{Alday} for a recent calculation for the dimension-2 case and \cite{Arutyunov:2002fh} for earlier work related to the dimension-3 case. Furthermore, in \cite{Cornalba1, Cornalba2,Cornalba3}, using Eikonal approximation methods pertaining to 2-2 scattering with spin-$\ell_m$ exchange in the gravity dual, the anomalous dimensions of large-$\ell$ and large-$n$ operators have been calculated.

In this paper we examine $\g(n,\ell)$ and OPE coefficients for general CFTs following \cite{anboot, Komargodski}. Our findings are consistent with AdS/CFT predictions \cite{Cornalba1, Cornalba2, Cornalba3} where it was found that for $\ell\gg n\gg1$, $\g(n,\ell)\propto -n^4/\ell^2$ while for $n\gg\ell\gg1$, $\g(n,\ell)\propto-n^3/\ell$ for graviton exchange dominance in the five dimensional bulk.

\vspace{0.7cm}
\noindent \textbf{Summary of the results:}\\
As we will summarize below, we can calculate the anomalous dimensions and OPE coefficients for the single tower of twist $2\Delta_\phi+2n$ operators with large spin-$\ell$ which contribute to one side of the bootstrap equation in an appropriate limit with the other side being dominated by certain minimal twist operators. In this paper we will focus on the case where the minimal twist $\tau_m=2$. One can consider various spins $\ell_m$ for these operators. We will present our findings for various spins separately; the case where different spins $\ell_m$ contribute together can be computed by adding up our results. We begin by summarizing the $\ell\gg n $ case. We note that, as was pointed out in \cite{anboot}, in this limit we do not need to have an explicit $1/N^2$ expansion parameter to make these claims. The $1/\ell^2$ suppression in both the anomalous dimensions and OPE coefficients does the job of a small expansion parameter\footnote{Strictly speaking we will need $\ell^2\gg n^4$ for this to hold. Otherwise we will assume that there is a small expansion parameter.}.

\begin{figure}
\begin{center}
\includegraphics[width=0.8\textwidth]{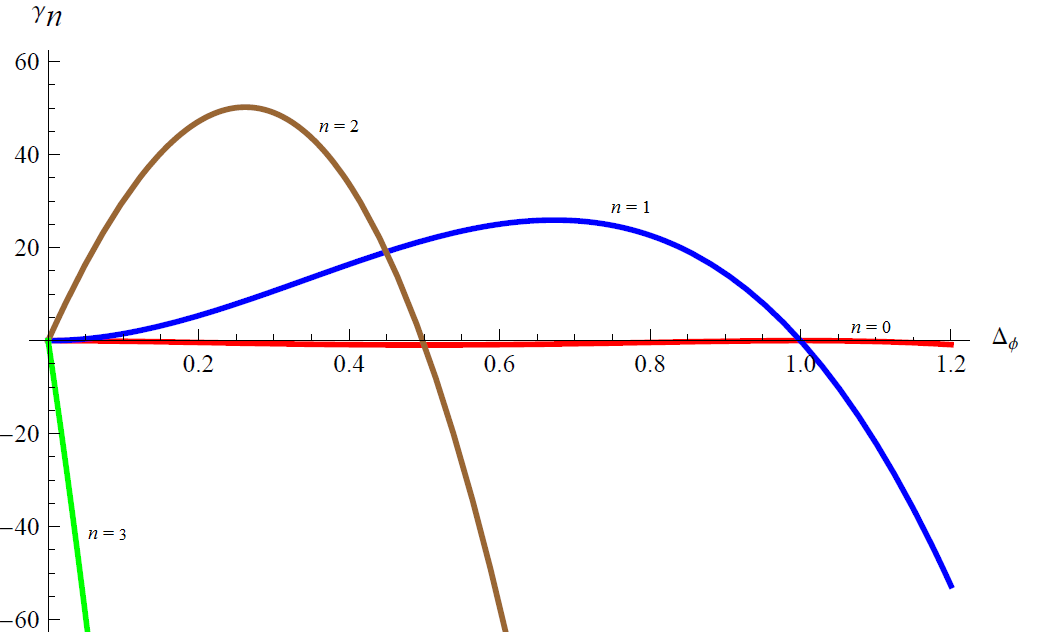}
\caption{The variation of the anomalous dimensions $\g_n$ with $\D_\phi$ showing that some of the anomalous dimensions become positive when $\D_\phi<(d-2)/2$.}
\label{fig:gnplot1}
\end{center}
\end{figure}

For the dominant $\tau_m=2, \ell_m=0$ contribution, the anomalous dimension becomes independent of $n$ and is given by,
\be
\g(n,\ell)=-\frac{P_m(\D_\phi-1)^2}{2\ell^2}\,.
\ee
while the correction to the OPE coefficient can be shown to approximate to,
\be\label{cn}
\mc_n=\frac{1}{\tq_{\D_\phi,n}}\partial_n(\tq_{\D_\phi,n}\g_n)\,,
\ee
in the large $n$ limit similar to the observation made in \cite{Polchinski}. The coefficient $\tq_{\D_\phi,n}$ is related to the MFT coefficients as shown in \eqref{tq} later. Here $P_m$ is related to the OPE coefficient corresponding to the $\tau_m=2, \ell_m=0$ operator.
For the dominant $\tau_m=2=\ell_m$ contribution, the anomalous dimension is given by,
\be
\g(n,\ell)=\frac{\g_n}{\ell^2},
\ee
where,
\begin{align}\label{gspin2intro}
\begin{split}
\g_n=&-\frac{15P_m}{\D_\phi^2}[6n^4+12n^3(2\D_\phi-3)+6n^2(11-14\D_\phi+5\D_\phi^2)\\
&+6n(2\D_\phi-3)(\D_\phi^2-2\D_\phi+2)+\D_\phi^2(\D_\phi-1)^2]\,.
\end{split}
\end{align}
Using the standard normalization (see \cite{Komargodski}), $P_m=2\D_\phi^2/(45N^2)$ and hence $P_m/\D_\phi^2$ becomes independent of $\D_\phi$. Thus for $n\gg 1$, $\g(n,\ell) N^2\approx-4n^4/\ell^2$, independent of $\D_\phi$. The coefficients $\g_n$ are negative for arbitrary $n$  and $\D_\phi \geq 1$. Interestingly (as shown in figure \eqref{fig:gnplot1}) some $\g_n$'s -- $n=1,2$--can become positive if $0<\D_\phi <1$, i.e., for $\D_\phi$ violating the unitarity bound. To make a connection between the unitarity bound and the sign of the anomalous dimension, having the exact analytic expression above was crucial. For example, without the exact expression it would not be possible to conclude that all anomalous dimensions above $n=2$ will be negative. Further, without such a formula, it would not be possible to infer the universality, i.e., independence of $\Delta_\phi$ in the $\ell \gg n\gg 1$ regime. Let us emphasise two points. First, that while the Eikonal approximation methods in AdS/CFT agree with our general formula above in this limit, the subleading terms in $n$ are in fact a prediction from bootstrap which will be interesting to verify using a gravity calculation. Second, the fact that $1/\ell$ plays the proxy for a small expansion parameter makes it clear that this result is valid not just for a large $N$ theory but for any theory satisfying the minimal set of assumptions about the spectrum mentioned above.

%
%
For general $\ell_m$ we find that the anomalous dimension behaves like 
\begin{equation}
\gamma(n,\ell) \propto -\frac{n^{2\ell_m}}{\ell^2}\,,
\end{equation}
for large $n$. The proportionality constant is related to the corresponding OPE coefficient. Even for this case, the anomalous dimensions are all negative for $\D_\phi$ respecting the unitarity bound and can be positive otherwise. Thus there appears to be an interesting correlation between CFT unitarity and bulk causality (in the sense that the sign of the anomalous dimension is correlated with the bulk Shapiro time delay \cite{Camanho:2014apa}). 
 
Let us make some observations. If we assume that  $\ell_m \leq 2$ as in \cite{anboot}, our results suggest that since the $\D_\phi$ dependence drops out in $\g_n$ for $n\gg1$, the findings are universal for any 4d CFT with a scalar of conformal dimension $\D_\phi$ and where in the $\ell\gg 1$ limit the spectrum is populated with a single tower of operators with twists $2\D_\phi+2n$ separated by a twist gap from other operators. 
The explicit results given in \cite{Alday,Arutyunov:2002fh} are indeed consistent with the universal form of our result at large $n$. Furthermore our result is consistent with the AdS/CFT calculations in the Eikonal approximation. This gives credence to our finding that in the limit $\ell\gg n\gg1$ the anomalous dimensions and the OPE  coefficients for the $\ell_m=2$ exchange indeed take on a universal form.

We will further extract the subleading $1/\ell^3$ correction to the anomalous dimension for stress tensor exchange dominance and show that in the limit $\ell\gg n\gg 1$, the result is universal as well. For this we will provide a systematic way to compute the corrections to the conformal blocks starting with the differential equation.

Our paper is organized as follows: we start with the review of the analytical bootstrap methods used in \cite{anboot,Komargodski} in section \eqref{sec2}. In section \eqref{sec3} we apply these methods in the limit when the spin is much larger than the twist, to cases where the \textit{lhs} of the bootstrap equation is dominated by either the twist-$2$, spin-$2$ operator exchange or a twist-$2$ scalar operator exchange.  In \eqref{subl} as a further extension, we consider the subleading terms in the $1/\ell$ expansion and compare with known results. In section \eqref{sec42} we compare our results with the ones from AdS/CFT. Specifically we find that our results are in agreement with the results in \cite{Cornalba1,Cornalba2,Cornalba3} in both the limits. We end the paper with a brief discussion of open questions in \eqref{sec6}. Certain useful relations and formulae used for \eqref{sec2} are discussed in appendices \eqref{secA} and \eqref{secB}. In appendix \eqref{secC} we give a brief detail of the $n$ dependence of the coefficients $\g_n$ for $\ell_m>2$ cases. In appendix \eqref{secD} we consider the differential equations which will lead to the extensions of the large $\ell$ results in the subleading orders in $\ell$ in \eqref{subl}. In appendix \eqref{sec5} we discuss the behaviour of the corrections to the OPE coefficients $\mc_n$ for $\ell\gg n$ limit where we show that asymptotically ( for large $n$), the coefficients $\mc_n$ approach the relation \eqref{cn} while at low $n$ there are deviations.  Finally in appendices \eqref{sec4}  and \eqref{secE} we address the other limit where the twist is much larger than the spin. These aim to provide an unified approach to handle both the limits ($\ell\gg n$ and $n\gg\ell$) using a saddle point analysis and our findings there are preliminary.

\section{Review of the analytical approach}\label{sec2}
We begin by reviewing the key results of \cite{anboot} (see also \cite{Komargodski}) which will help us set the notation as well. 
Consider the scalar 4-point correlation function $\langle \phi(x_1)\phi(x_2)\phi(x_3)\phi(x_4) \rangle$. In an arbitrary conformal field theory, we have a $12\rightarrow 34$ OPE decomposition (s-channel) given by,
\be
\langle \phi(x_1)\phi(x_2)\phi(x_3)\phi(x_4)\rangle =\frac{1}{x_{12}^2 x_{34}^2}\sum_{\mathcal{O}}P_{\mathcal{O}}\ g_{\tau_{\mathcal{O}},\ell_{\mathcal{O}}}(u,v)\,.
\ee
Here we have used the notation $x_{ij}=x_i-x_j$. The variables $u$ and $v$  are the conformal cross ratios defined by,
\be
u=\frac{x_{12}^2 x_{34}^2}{x_{24}^2 x_{13}^2}\text{,}\hspace{0.5cm}\text{and }\hspace{0.5cm}v=\frac{x_{14}^2 x_{23}^2}{x_{24}^2 x_{13}^2}\,,
\ee
The functions $g_{\tau_{\mathcal{O}},\ell_{\mathcal{O}}}(u,v)$ are called conformal blocks or conformal partial waves \cite{Dolan}, and they depend on the spin $\ell_{\mathcal{O}}$ and twist $\tau_{\mathcal{O}}$ of the operators $\mathcal{O}$ appearing in the OPE spectrum. The twist is given by $\tau_{\mathcal{O}}=\D_{\mathcal{O}}-\ell_{\mathcal{O}}$, where $\D_{\mathcal{O}}$ is the conformal dimension of $\mathcal{O}$. $P_{\mathcal{O}}$ is a positive quantity related to the OPE coefficient. The sum goes over all the twists $\tau$ and spins $\ell$ that characterize the  operators. 

The 4-point function will also have a decomposition in the $14\rightarrow 23$ channel (t-channel), and equating the two channels we will have the bootstrap equation,
\be
1+\sum_{\ta,\ell}P_{\ta,\ell}\ g_{\ta,\ell}(u,v)=\left(\frac{u}{v}\right)^{\D_\phi}\left(1+\sum_{\ta,\ell}P_{\ta,\ell}\ g_{\ta,\ell}(v,u)\right)\,.
\ee
We will work in the limit $u \ll v < 1$. In this limit the leading term on the \textit{lhs} is the 1. However on the \textit{rhs} $g_{\tau,\ell}$ has no negative power of $u$ in the small $u$ limit and all terms are vanishingly small. So we cannot reproduce the leading 1 from the \textit{rhs} from a finite number of terms. In mean field theory it was shown \cite{anboot} that the large $\ell$ operators produce the leading term. For a general CFT, the authors of \cite{anboot} argued that in order to satisfy the leading behavior,
\be\label{leadinggencft}
1 \approx \left( \frac{u}{v} \right)^{\D_\phi}\sum_{\ta,\ell}P_{\ta,\ell}g_{\ta,\ell}(v,u)\,,
\ee
the twists $\tau$ must have the same pattern as in MFT. To show this we have to look at the large $\ell$ and small $u$ limit of the conformal blocks,
\begin{align}\label{large l block}
g_{\tau,\ell}(v,u)&=k_{2\ell}(1-z)v^{\tau/2}F^{(d)}(\tau,v)\,,\hspace{1cm}(\text{when } |u|\ll 1\text{ and }\ell \gg 1)\nonumber\\k_{\beta}(x)&=x^{\beta/2} {_2}F_1(\beta/2,\beta/2,\beta,x)\,.
\end{align}
Here $z$ is defined by $u=z \bar{z}\,, v=(1-z)(1-\bar{z})$; and  $F^{(d)}(\tau,v)$ is a positive and analytic function near $v=0$ whose exact expression is not necessary for the discussion. We derive the above result later in this section. For now, we just use this to rewrite (\ref{leadinggencft}),
\be\label{varseparation}
1 \approx \sum_{\tau}\left(\lim\limits_{z\to 0} z^{\D_{\phi}}\sum_{\ell} P_{\tau,\ell}k_{2\ell}(1-z)\right)v^{\tau/2-\D_\phi}(1-v)^{\D_\phi}F^{(d)}(\tau,v)\,.
\ee
The term in brackets are independent of $z$ and $\ell$ after taking the limit and doing the sum (over $\ell$). Then what is left is just a function of $\tau$ with a sum over $\tau$. The function $F^{(d)}(\ta,v)$ around small $v$ begins with a constant. Thus we must have $\ta/2=\D_\phi$ in the spectrum. Next since $F^{(d)}(\ta,v)$has terms with higher powers in $v$, we must have $\ta=2\D_\phi+2n$ for every integer $n$, to cancel these terms. This shows that there are operators with twists $\ta=2\D_\phi+2n$. Since these are operators in MFT, $P_{\ta,\ell}=P^{MFT}_{\ta,\ell}$ at leading order. We will now focus our attention on the subleading terms of the bootstrap equation.

The subleading corrections to the bootstrap equation are characterized by the anomalous dimension $\g(n,\ell)$ and corrected OPE coefficients $\mc_n$. We will assume that for each $\ell$ there is a single operator having twist $\tau \approx 2\D_{\phi}+2n$. The bootstrap equation takes the form\footnote{Our conventions for $P_m$ differ from \cite{anboot} by a factor of 1/4.},
\be\label{largel}
1+\sum_{\ell_m} \frac{P_m}{4} u^{\ta_m/2} f_{\ta_m,\ell_m}(0,v)\approx \sum_{\ta,\ell}P_{\ta,\ell}v^{\ta/2-\D_\phi}u^{\D_\phi}f_{\ta,\ell}(v,u)\,,
\ee 
which is valid upto subleading corrections in $u$ as $u\rightarrow 0$. Note that the \textit{lhs} demands the existence of an operator of minimal twist $\ta_m=\D_m-\ell_m$ which is non-zero. We set $u=z(1-v)+O(z^2)$ and consider $u\rightarrow0$ to be $z\rightarrow0$. The explicit form of the function $f_{\ta_m,\ell_m}(v)$ is given by,
\be\label{largel1}
f_{\ta_m,\ell_m}(v)=\frac{\G(\ta_m+2\ell_m)}{\G\bigg(\ta_m+\frac{\ell_m}{2}\bigg)^2}(1-v)^{\ell_m}\sum_{n=0}^\infty\bigg(\frac{(\ta_m+2\ell_m)_n}{n!}\bigg)^2 v^n\bigg[2[\psi(n+1)-\psi\bigg(\frac{\ta_m}{2}+\ell_m+n\bigg)]-\log v \bigg].
\ee 
Later we will set $\ta_m=2$ because we are particularly interested in the twist 2 primary operator or the stress tensor in the theory.


Let us now focus on the \textit{rhs} where we have an infinite sum over all twists and spins.  In the limit $\ell\gg n\gg 1$ we can simplify the \textit{rhs} considerably. Note that we will be working in $d=4$ since in the $d=2$ case there is no minimal twist operator with a twist gap from the identity operator ($\tau_{min}^{d=2}=0$). To proceed we first need to find the behaviour of the conformal blocks in the above limit ( in other words $\ta_m=2$) and when $|u|\ll |v|< 1$. With  $u=z(1-v)+O(z^2)$ since $\bz=(1-v)+O(z)$, we can form a small $z$ expansion around $z=0$ and then a small $v$ expansion. To find the anomalous dimension $\g(n,\ell)$ for each $\ell$ we need to match the coefficients of the terms $v^n\log v$ on both sides of \eqref{largel}. Considering $\ta(n,\ell)=2\D_\phi+2n+\g(n,\ell)$, we can see that the $\log v$ arises from the next to the leading term in the perturbative expansion around small $v$ given by,
\be
v^{\ta(n,\ell)/2-\D_\phi}\rightarrow \frac{\g(n,\ell)}{2}v^n\log v\,.
\ee
The MFT coefficients take the following form in the $\ell \gg n$ limit,
\be
P_{2\D_\phi+2n,\ell}\overset{\ell\gg1}{\approx}q_{\D_\phi,n}\frac{\sqrt{\pi}}{2^{2\D_\phi+2n+2\ell}}\ell^{2\D_\phi-3/2}\,,
\ee
where the coefficient $q_{\D_\phi,n}$ is given by,
\be
q_{\D_\phi,n}=\frac{8}{\G(\D_\phi)^2}\frac{(1-d/2+\D_\phi)_n^2}{n!(1-d+n+2\D_\phi)_n}.
\ee
Here $(a)_b=\G(a+b)/\G(a)$ is the Pochhammer symbol. We will also use another notation for convenience in the later part of the work,
\be\label{tq}
\tq_{\D_\phi,n}=2^{-2\D_\phi-2n}q_{\D_\phi,n}\,.
\ee
The $d=4$ crossed conformal blocks are given by 
\be\label{cbll}
g_{\ta,\ell}(v,u)=\frac{(1-z)(1-\bz)}{\bz-z}[k_{2\ell+\ta}(1-z)k_{\ta-2}(1-\bz)-k_{2\ell+\ta}(1-\bz)k_{\ta-2}(1-z)]\,,
\ee
where we have already defined $k_{\beta}(x)$ in \eqref{large l block}. As already mentioned, in the large $\ell$ limit, the conformal blocks simplify to give (\ref{large l block}). For $\ell \gg n$ we can decompose $k_{2\ell+\ta}(1-z)$ even further to get,
\be
k_{2\ell+\ta}(1-z)\overset{\ell\to \infty}{\approx} \frac{2^{\ta+2\ell-1}\ell^{1/2}}{\sqrt{\pi}} K_0(2\ell\sqrt{z})\,.
\ee
We will also need the expression for $F^{(d)}(\tau,v)$. In $d=4$ we have, 
\be
F^{(4)}=\frac{2^{\ta}}{1-v} \ {}_2F_1\bigg[\frac{\ta}{2}-1,\frac{\ta}{2}-1,\ta-2,v\bigg]\,.
\ee
With this, the entire ($ \log v$ dependent part of) \textit{rhs} of \eqref{largel} in the limit $\ell\gg n$ can be organized into the following form,
\begin{align}\label{rhs}
\begin{split}
\sum_{\ta,\ell}P_{\ta,\ell}v^{\ta/2-\D_\phi}u^{\D_\phi}f_{\ta,\ell}(v,u)=&\sum_{n=0,\ell=\ell_0}^{\infty} \frac{q_{\D_\phi,n}}{2} \ell^{2\D_\phi-\frac{3}{2}}\bigg[\frac{\g(n,\ell)}{2}\bigg]v^n \log v  \ \ell^{1/2} K_0(2\ell\sqrt{z})z^{\D_\phi}\\
&(1-v)^{\D_\phi-1}{}_2F_1(\D_\phi+n-1,\D_\phi+n-1,2\D_\phi+2n-2;v)\,.
\end{split}
\end{align}

Now the overall factor of $u^{\D_\phi}$ sitting on the \textit{rhs} of (\ref{largel}) is translated into an overall factor of $z^{\D_\phi}(1-v)^{\D_\phi}$. We assume that the anomalous dimension has the form $\g(n,\ell)=\g_n/\ell^\a$. Now in the large $\ell$ limit we can convert the sum over $\ell$ in (\ref{rhs}) into an integral given by,
\be
\int_{\ell_0}^\infty d\ell \  \ell^{-1-\a+2\D_{\phi}}z^{\D_{\phi}}K_0(2\ell\sqrt{z})\approx \frac{z^{\a/2}}{4}\G^2\bigg(\D_\phi-\frac{\a}{2}\bigg)+O(z^{\D_\phi} \log z)\,.
\ee
In order to do this integral, it is convenient to use an upper cutoff $L$. The integral works out to be in terms of regularized Hypergeometric functions. By expanding the result assuming $L \sqrt{z}\gg 1$ and $\ell_0 \sqrt{z} \ll 1$ we get the leading and subleading terms in the above equation. For $\D_\phi >1$, the $O(z^\D_\phi \log z)$ terms can be ignored. 
 This reproduces\footnote{Note that for $\D_\phi=1$ and $\tau_m=2$, this does not work as the Gamma function blows up. This is presumably indicative of a $\log \ell$ scaling for the operators \cite{Bissi} in this case.} the factor of $z^\frac{\tau_m}{2}$ exactly if $\a=\ta_m$. If we take the minimal nonzero twist to be $\ta_m=2$, the anomalous dimension behaves as,
\be\label{anomdenom}
\gamma(n,\ell)=\frac{\gamma_n}{\ell^{2}}\,.
\ee
Once again the interested reader should refer to \cite{anboot, Komargodski} for the mathematical details of the above algebra and approximations.  In the next section, we demonstrate how the expression for $\g_n$ can be given in terms of an exact sum for all $n$. This sum enables us to extract the exact behaviour of the anomalous dimensions for all $n$ when $\ell \gg n$. Later in appendix \eqref{sec4} we have also considered anomalous dimensions for the other limit $\ell\gg n\gg1$.

\section{The $\ell \gg n$ case }\label{sec3}
We begin by determining $\g_n$ appearing in (\ref{anomdenom}) in the limit $\ell \gg n$. To get $\g_n$, we have to match the power of $v^n\log v$ on both sides of (\ref{largel}). To do that we take the $(1-v)^{\D_\phi-1}$ of (\ref{rhs}) to the \textit{lhs} of (\ref{largel}) and expand $(1-v)^{\ell_m+\tau_m/2-\D_\phi+1}$ in powers of $v$. Thus the \textit{lhs} of \eqref{largel} becomes,
\be
-(1-v)^{\ta_m/2+\ell_m+1-\D_\phi}\frac{P_m}{4}\frac{\G(2\ell_m+\ta_m)}{\G(\ell_m+\ta_m/2)^2}\sum_{n=0}^\infty \bigg(\frac{(\ta_m/2+\ell_m)_n}{n!}\bigg)^2v^n\log v\,,
\ee
where $(a)_b$ is the Pochhammer symbol. Expanding the term $(1-v)^{\ell_m+\tau_m/2-\D_\phi+1}$, we get,
\be
(1-v)^{\ell_m+\tau_m/2-\D_\phi+1}=\sum_{\a=0}^\infty (-1)^k\frac{b!}{\a!(b-\a)!}v^\a \hspace{0.5cm}\text{where}\hspace{0.5cm}b=\ell_m+\frac{\tau_m}{2}+1-\Delta_\phi\,.
\ee
Now set $n+\a=k$ whereby the \textit{lhs} can be arranged as $\sum_{n=0}^\infty L_n v^n \log v$ where to find $L_k$ we need to perform the $\a$ sum explicitly. 

\begin{figure}
\begin{center}
\includegraphics[width=0.8\textwidth]{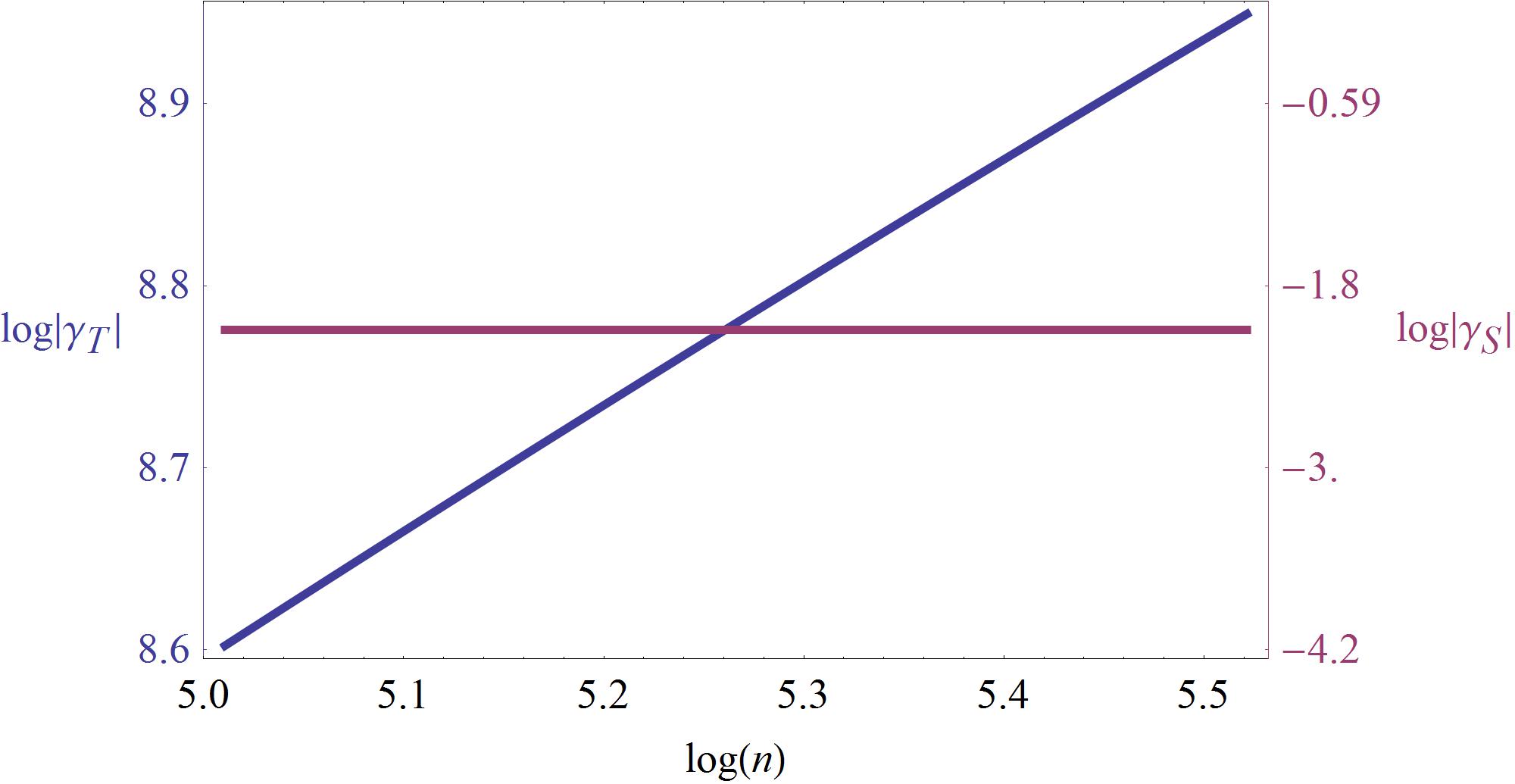}
\caption{$\log |\g_n|$ vs. $\log n$ plot showing the dependence of $\g_n$ on $n$ for $n\gg1$. $\g_T$ is the anomalous dimension for the spin-$2$ operator exchange and $\g_S$ for the scalar operator exchange. The slope of the blue straight line for spin-$2$ exchange is $3.998$ while the red line denotes the scalar exchange for which $\g_n$ is constant for all $n$. We have used $\D_\phi=2$ in the above plots.}
\label{fig:gnplot}
\end{center}
\end{figure}

This gives, the coefficient of $v^n \log v$ to be,
\be
L_n=-4P_m\frac{\G(\ta_m+2\ell_m)}{\G\bigg(\frac{\ta_m}{2}+\ell_m\bigg)^2}\sum_{\a=0}^\infty (-1)^\a\bigg(\frac{(\ta_m/2+\ell_m)_{(n-\a)}}{(n-\a)!}\bigg)^2 \frac{b!}{(b-\a)!\a!}\,,
\ee
where we have multiplied the \textit{lhs} of \eqref{largel} with an overall numerical factor of $16$ coming from the \textit{rhs} of \eqref{largel}.
This finite sum is given by,
\be
L_n=-\frac{4P_m\Gamma \left(2 \ell_m+\tau _m\right)\Gamma \left(\text{  }n+\ell_m+\frac{\tau _m}{2}\right){}^2 \ {}_3F_2\left(
\begin{array}{c}
 -n,-n,-1-\ell_m+\D_\phi -\frac{\tau _m}{2} \\
 1-n-\ell_m-\frac{\tau _m}{2},1-n-\ell_m-\frac{\tau _m}{2}
\end{array}
,1\right)}{\Gamma (1+n)^2 \Gamma \left(\ell_m+\frac{\tau _m}{2}\right)^4 }\,.
\ee
To get the same coefficient of $v^n \log v$ on the \textit{rhs} of \eqref{largel}, we expand the hypergeometric function in powers of $v$ given by
\be
_2F_1(\ta/2-1,\ta/2-1,\ta-2,v)=\sum_{\a=0}^\infty \frac{(\ta/2-1)_\a^2}{(\ta-2)_\a\ \a!} v^\a\,,
\ee
where $(a)_b$ is the Pochhammer symbol given by $(a)_b=\G(a+b)/\G(a)$. On the \textit{rhs} we have two infinite sums $\Sigma_{k=0}^\infty\Sigma_{\a=0}^\infty f_{\a,k} v^{k+\a}$. To put the \textit{rhs} in the form $\Sigma_{n=0}^\infty R_n v^n$ we will regroup the terms in the double sum in increasing powers of $v^n$. This is achieved by setting $k+\a=n$ where $\a$ runs from $0$ to $n$ giving,
\be
\textit{rhs}=\sum_{n=0}^\infty R_n \ v^n \log v\,,
\ee
where, the coefficients $R_k$ can be written as
\be
R_n=\G(\D_\phi-\frac{\tau_m}{2})^2\sum_{\a=0}^{n} q_{\D_\phi,n-\a}\g_{n-\a}  \left(\frac{\left(\frac{\tau }{2}-1\right)_{n-\a}^2}{(n-\a)!(\tau -2)_{n-\a}}\right)\,,
\ee
where the extra factor of $\frac{1}{2}$ comes from the normalization $2^{2\ell+\ta-1}$ when we consider the large $\ell$ approximation of the conformal blocks. Equating the coefficients $R_n=L_n$ we can find the corresponding coefficients $\g_n$. Thus, in principle, we would know $\g_n$ if we know $\g_k$ for all $k\le n-1$. In figure \eqref{fig:gnplot} we have plotted the $\log \g_n$ vs. $\log n$ for a twist-$2$ scalar and a twist-$2$ and spin-$2$ field. 

We find that the slope of the curve for the twist-$2$, spin-$2$ exchange is $\approx4$ while that for the twist-$2$ scalar is a constant. So $\g_n \sim n^4$ for large values of $n$ for spin-$2$ field. To show this behavior explicitly, we notice that $\g_n$ can be written as an exact sum over the coefficients $R_m$ appearing on the \textit{lhs}. This formula can be guessed by looking at the first few $\g_n$s. We give the form of the first few $\g_n$s. These take the form\footnote{We will assume $\D_\phi>1$. See footnote 4.},
\begin{align}
\begin{split}
\g_0&=\frac{(\D_\phi-1)^2}{8}L_0\,,\\
\g_1&=-\frac{(\D_\phi-1)^2}{8}L_0+\frac{\D_\phi-1}{4}L_1\,,\\
\g_2&=\frac{(\D_\phi-1)^2}{8}L_0-\frac{2\D_\phi-1}{4}L_1+\frac{2\D_\phi-1}{2\D_\phi}L_2\, \ \text{etc.}
\end{split}
\end{align}
We observe that the above terms follow a definite pattern which can be written as,
\be\label{gamma sum}
\g_n =\sum_{m=0}^{n} a_{n,m}\hspace{1cm}\text{with}\hspace{1cm}a_{n,m}=c_{n,m}L_m\,.
\ee
where for general $\ta_m$ and $\ell_m$ the coefficients $c_{n,m}$ are given by,
\be\label{cnm}
c_{n,m}=\frac{1}{8}\left(\frac{\Gamma \left(\Delta _{\phi }\right)}{\Gamma \left(\Delta _{\phi }+m-1\right)}\right)^2 \frac{\left(2\Delta _{\phi }+n-3\right)_m (-1)^{n+m} n!}{(n-m)!}\bigg(\frac{\G(\D_\phi-1)}{\G(\D_\phi-\ta_m/2)}\bigg)^2\,.
\ee
We have checked the analytic expression for the coefficients $\g_n$ agrees with the solutions of $\g_n$ found from solving the equations $R_k=L_k$ order by order for arbitrary values of $n$. 


\subsection{Case I: $\ta_m=2$, $\ell_m=0$ }\label{sec31}
We now consider the case where the \textit{lhs} of \eqref{largel} is dominated by the exchange of a twist-$2$ scalar operator.  For this case 
\begin{align}
\begin{split}
 _3F_2\left[
\begin{array}{c}
 -m,-m,-2+\Delta_\phi  \\
 -m,-m
\end{array}
,1\right]&=\sum_{k=0}^m\frac{\G(k+\D_\phi-2)}{\G(\D_\phi-2)k!}=\frac{\G(\D_\phi+m-1)}{\G(m+1)\G(\D_\phi-1)}\, .
\end{split}
\end{align}
The coefficients $a_{n,m}$ can thus be written as,
\be
a_{n,m}=-\frac{P_m}{2}\frac{(-1)^{m+n}(\D_\phi-1)\G(n+1)\G(\D_\phi)\G(2\D_\phi+m+n-3)}{\G(m+1)\G(n+1-m)\G(\D_\phi+m-1)\G(2\D_\phi+n-3)}\,.
\ee
We sum over the coefficients $a_{n,m}$ to get,
\be
\g_n=\sum_{m=0}^n a_{n,m}=-\frac{P_m}{2}(\D_\phi-1)^2\,.
\ee
Note that the coefficients $\g_n$ appearing in the expression for the anomalous dimension become independent of $n$ in this case. The details can be found in appendices \eqref{secA} and \eqref{secB}.

\subsection{ Case II: $\ta_m=2$, $\ell_m=2$}\label{sec32}

Here we consider the case where the \textit{lhs} of \eqref{largel} is dominated by the exchange of a twist-$2$ and spin-$2$ operator exchange. In the language of AdS/CFT, the particle is a graviton that dominates the scattering amplitude in the Eikonal limit \cite{Cornalba1, Cornalba2, Cornalba3}. As in the previous case the anomalous dimension goes as $\sim 1/\ell^2$ for large spin in the \textit{rhs} of \eqref{largel}. Performing the $\ell$ integration we are left with a single sum on the \textit{rhs} from which we can determine the coefficients $\g_n$ as a function of $n$. Using relation \eqref{gamma sum} we can evaluate the coefficients $L_m$ for the case when $\ta_m=2$ and $\ell_m=2$ respectively which we proceed to show below. We defer the details of the calculation to the appendix and present here with only the final results. First we write
\begin{align}
\begin{split}
 _3F_2\left[
\begin{array}{c}
 -m,-m,-4+\Delta_\phi  \\
 -2-m,-2-m
\end{array}
,1\right]&=\sum_{k=0}^m \frac{(m+1-k)^2(m+2-k)^2\G(\D_\phi-4+k)}{(m+1)^2(m+2)^2\G(k+1)\G(\D_\phi-4)}\\
&=\frac{4(6m^2+6m(\D_\phi-1)+\D_\phi(\D_\phi-1))\G(m+\D_\phi-1)}{(m+1)(m+2)\G(m+3)\G(\D_\phi+1)}\,.
\end{split}
\end{align}
The combined coefficients $a_{n,m}$, after putting in the proper normalizations, can be written as,
\begin{align}
\begin{split}
a_{n,m}&=-(-1)^{m+n}\frac{15 P_m}{\D_\phi}(6m^2+6m(\D_\phi-1)+\D_\phi(\D_\phi-1))\\
&\times\frac{\G(n+1)\G(\D_\phi)\G(2\D_\phi+m+n-3)}{\G(m+1)\G(n+1-m)\G(\D_\phi+m-1)\G(2\D_\phi+n-3)}\, .
\end{split}
\end{align}
We can now perform the summation, over the coefficients $a_{n,m}$ to get,
\begin{align}\label{gspin2}
\begin{split}
\g_n=\sum_{m=0}^n a_{n,m}=&-\frac{15P_m}{\D_\phi^2}[6n^4+\D_\phi^2(\D_\phi-1)^2+12n^3(2\D_\phi-3)+6n^2(11-14\D_\phi+5\D_\phi^2)\\
&+6n(2\D_\phi-3)(\D_\phi^2-2\D_\phi+2)]\,.
\end{split}
\end{align}
The above formula negative and monotonic for all values of $n$ and $\D_\phi>1$ (see appendices \eqref{secA} and \eqref{secB} for details). Until this point we did not need the explicit form of the coefficient $P_m$ but we can choose the conventions\cite{Komargodski}. $P_m$ for any general $d$ is given by
\be
P_m=\frac{d^2}{(d-1)^2}\frac{\D_\phi^2}{C_T}\,.
\ee
This result follows from the conformal Ward identity\footnote{We thank Joao Penedones for reminding us of this fact.}; as a consequence the $\D_\phi$ independence of the $n^4$ term in the anomalous dimension is a general result. For our case we put $d=4$ and $C_T=40N^2$, which correspond to the AdS$_5$/CFT$_4$ normalization and  where $C_T$ is the central charge. Putting all these together, we get,
$
P_m=\frac{2}{45 N^2}\D_\phi^2.
$
Note that the $n^4$ term in $\gamma_n$ becomes independent of $\D_\phi$ using this convention. Thus when $n$ is large, the result is independent of $\D_\phi$ and hence universal.

\subsection{Comment on the ${\mathcal{N}}=4$ result}\label{sec33}
In \cite{Alday}, the authors showed that for dimension-2 half-BPS multiplet the anomalous dimension in ${\mathcal{N}}=4$ SYM, for $\D_\phi=2$, has the form,
\be
\g(n,\ell) N^2=-\frac{4(n+1)(n+2)(n+3)(n+4)}{(\ell+1)(\ell+6+2n)}\,.
\ee
To compare this with our result (\ref{gspin2}) we put $P_m=2/(45N^2)\D_\phi^2$ (See for eg. \cite{Komargodski}), and set $\D_\phi=4.$ This gives,
\be
\g(n,\ell) N^2\approx -\frac{4(n+1)(n+2)(n+3)(n+4)}{\ell^2}\,.
\ee
for large values of $\ell$. Quite curiously this form matches with the supergravity result, for large spin and finite $n$. The reason for this agreement is not clear to us although \cite{Alday} made a similar observation that the extra solutions to the bootstrap equation they find (for $\D_\phi=2$) match exactly with the solutions in \cite{Polchinski} for $\D_\phi=4$.

\section{Subleading terms at large spin and large twist}\label{subl}

So far we have concentrated on the leading $n$ dependence in the anomalous dimensions for large spin operators, when the \textit{lhs} of the bootstrap equation is dominated by the stress tensor exchange. Let us now see how to derive the first subleading term with the stress tensor exchange. In \cite{Alday1}, this problem for leading twist was considered. By considering large twists we will extract universal results. It turns out that just keeping the leading $\ell$ dependence of the \textit{rhs} is not sufficient anymore. We will assume a $1/N$ expansion so that we can use the $1/\ell$ corrections in $P^{MFT}$--without the large $N$ we would need to keep track of corrections to these coefficients as well. In $P^{MFT}$ the subleading corrections at large $\ell$ take the form,
\be
P^{MFT}=\frac{\sqrt{\pi}\ell^{2\D_\phi-3/2}}{2^{2\D_\phi+2\ell}}\bigg[q_{\D_\phi,n}+\frac{1}{\ell}r_{\D_\phi,n}\bigg]\,,
\ee
where the coefficients $q_{\D_\phi,n}$ and $r_{\D_\phi,n}$ are given by,
\begin{align}
q_{\D_\phi,n}&=\frac{8(\D_\phi-1)_n{}^2}{(2\D_\phi+n-3)_n\G(n+1)\G(\D_\phi)^2}\,,\nonumber\\ 
r_{\D_\phi,n}&=\frac{(\D_\phi-1)_n{}^2}{(2\D_\phi+n-3)_n\G(n+1)\G(\D_\phi)^2}[5-20\D_\phi+16\D_\phi^2+4n(4\D_\phi-3)]\,.
\end{align} 
The conformal blocks in the crossed channel, in the large $\ell$ and $u\rightarrow0$ limit is,
\be
g_{\ta,\ell}(v,u)\overset{\ell\gg1}{\underset{u\ll1}\approx}k_{2\ell+\ta}(1-z)v^{\ta/2}F^{(d)}(\ta,v)\,,
\ee
where we have neglected the second term in above expression since even at subleading order in $\ell$, those terms will be exponentially suppressed. We can write $k_{2\ell+\ta}(1-z)$ as follows,
\be
k_{2\ell+\ta}(1-z)=\frac{\G(2\ell+\ta)}{\G(\ta/2+\ell)^2}\int_0^1 \frac{dt}{t(1-t)}\bigg(\frac{(1-z)t(1-t)}{1-t(1-z)}\bigg)^{\ell+\ta/2}\,.
\ee
 As we show in the appendix \eqref{secD}, for $\ell\gg n \gg 1$,
\be
k_{2\ell+\ta}(1-z)=\frac{\G(2\ell+\ta)}{\G(\ell+\frac{\ta}{2})^2}K_0((2\ell+\ta)\sqrt{z})+O(z)\,.
\ee
We can further approximate the $K_0$ function in the limit of small $\ta/\ell$ upto first order and similarly for the $\G$-functions. The relevant part of the conformal block in the crossed channel takes the form,
\be\label{subu}
k_{2\ell+\ta}(1-z)=\frac{2^{2\ell+\ta-1}}{\sqrt{\pi}}\ell^{1/2}\bigg[1+\frac{2\ta-1}{8\ell}\bigg][K_0(2\ell\sqrt{z})-\sqrt{z}\ta K_1(2\ell\sqrt{z})]\,.
\ee
Upto this order is sufficient for the calculation of the first subleading order in $z$ after $z^{\ta_m/2}$. Let us assume that the anomalous dimensions can be expanded in the form,
\be
\g(n,\ell)=\frac{\g^0_n}{\ell^{\ta_m}}+\frac{\g^1_n}{\ell^{\ta_m+1}}+\frac{\g_n^2}{\ell^{\ta_m+2}}+\dots\,.
\ee
>From the first subleading correction we should be able to determine $\g^1_n$. We already know the coefficient $\g_n^0$ as an exact function of $n$ and $\D_\phi$. Similar to the leading order case, we can perform the large $\ell$ summation (as an integral) and then evaluate the coefficients of the subleading powers of $z$ resulting from these extra terms. On the \textit{lhs}, the subleading powers of $z$ take the form,
\be
1+\frac{1}{4}P_m z^{\ta_m/2} g_{\ta_m,\ell_m}(u,v)=1+z^{\ta_m/2} f_1(v)\log v+z^{\ta_m/2+1} f_2(v)\log v+O(z^{\ta_m/2+2})\,.
\ee
Thus on the \textit{lhs} only integer powers of $z$ are there in the subleading pieces. Whereas on the \textit{rhs}, after the large $\ell$ integral, the first subleading power after the leading term in $z^{\ta_m/2}$ begins with $z^{(\ta_m+1)/2}$ with the coefficient,
\begin{align}
z^{(\ta_m+1)/2}\sum_n [(1-2\D_\phi+(2n+2\D_\phi)\ta_m)\g_n^0+2\g_n^1]&\frac{(\D_\phi-1)_n{}^2\G(\D_\phi-\frac{1}{2}-\frac{\ta_m}{2})^2}{2\G(n+1)\G(\D_\phi)^2\G(2\D_\phi+n-3)_n}\nonumber\\
&\times v^n \log v \ F^{(d)}[2\D_\phi+2n,v]\,.
\end{align} 
Since there is no $z^{(\ta_m+1)/2}$ term on the \textit{lhs}, we must have for large $n$,
\be
\g_n^1=-n\ta_m\g_n^0\,.
\ee
Thus specializing to the case of $\ta_m=2$ (stress tensor), to the first subleading order in $\ell$ we have for the anomalous dimensions,
\be\label{subgn}
\g(n,\ell)=\g^0(n,\ell)\bigg(1-\frac{2n}{\ell}\bigg)\,.
\ee
where $\g^0(n,\ell)= -4n^4/(N^2\ell^2)$. The main result of  \cite{Alday1} is still consistent with this finding\footnote{We thank Fernando Alday for the following observation.}. For $n=0$ for large $\ell$ the Casimir is $j^2\approx \ell^2$. For general $n$ the Casimir will become $j^2\approx (\ell+n)^2$. The main conclusion of \cite{Alday1} is that only even powers of $j$ should appear in the large spin limit--this was explicitly shown for leading twists. In terms of $j$  (for $\tau_m=2$) we could have $1/j^2\sim 1/(\ell+n)^2$ or $1/(j^2-n^2)\sim 1/(\ell (\ell+2n))$ so that in the $j$ variable only even powers of $j$ appear--both these forms are compatible with the subleading term we have derived. We will find the latter behaviour in what follows which is also consistent with the results of \cite{Alday} for the supersymmetric ${\mathcal N}=4$ case.

\section{Comparison with AdS/CFT}\label{sec42}
AdS/CFT provides us with a formula for the anomalous dimensions in terms of the variables $\bar{h}=\D_\phi+n$, $ h=\bar{h}+\ell$. In the limit $h,\bar{h}\rightarrow \infty$, the form of the anomalous dimension is given by \cite{Cornalba1,Cornalba2,Cornalba3},
\be
\g_{h,\bar{h}}=-c \  2^{2\ell_m-2}(h\bh)^{\ell_m-1} \Pi(h,\bar{h})\,,
\ee
where $\ell_m$ is the spin of the minimal twist operator, $\Pi(h,\bar{h})$ is a particular function of $h,\bar{h}$. In $d=4$  the function $\Pi(h,\bh)$ is given by
\be
\Pi(h,\bh)=\frac{1}{2\pi}\frac{h^2}{h^2-\bh^2}\bigg(\frac{h}{\bh}\bigg)^{1-\D_m}\,,
\ee
where $\D_m$ is the dimension of the minimal twist operator. Using $\D_m=\ta_m+\ell_m$ for operators with minimal twist-$\ta_m$ and spin-$\ell_m$, the expression for the anomalous dimension in 4d becomes,
\be
\g_{h,\bar{h}}=-2^{2\ell_m-3}\frac{c}{\pi}\frac{\bar{h}^{-2+2\ell_m+\ta_m}h^{2-\ta_m}}{h^2-\bar{h}^2}\,.
\ee
Neglecting the factor of $\D_\phi$ when both $n,\ell\gg 1$ we can write the above formula in terms $n,\ell$ giving,
\be
\g(n,\ell)=-2^{2\ell_m-3}\frac{c}{\pi}\frac{n^{-2+2\ell_m+\ta_m}(n+\ell)^{2-\ta_m}}{\ell(2n+\ell)}\,.
\ee
 The functional dependence on $\ell,n$ is exactly what we found from the CFT analysis. For $\ta_m=2$, in the limit $\ell\gg n\gg 1$ we can see that the above formula reduces to $\g(n,\ell)=-(2^{2\ell_m-3}c/\pi) (n^{2\ell_m}/\ell^2)$ while in the opposite limit it gives,  $\g(n,\ell)=-(2^{2\ell_m-4}c/\pi) (n^{2\ell_m-1}/\ell)$, where $\ell_m$ is the spin of the minimal twist operator. Further for $\ell_m=2$, with $c=2\pi/N^2$ our results for the two limits match exactly with the above prediction from AdS/CFT for the graviton (stress tensor) exchange. Also for $\ell_m>2$ the $n$ and $\ell$ dependence of the above expression is the same as given by our analysis (see appendix \ref{secC}).

\section{Discussion}\label{sec6}
We conclude by listing some open problems.
\begin{itemize}
\item It will be nice to extend our results to other dimensions, especially odd dimensions where the conformal blocks are not known in closed form \footnote{We have recently done this in \cite{boot2}.}.

\item It will be interesting to understand if and how stringy modes can make the anomalous dimensions in the limit $n\gg \ell \gg 1$ small. We have made some preliminary observations about this limit in appendix F.

\item One could use the results of our paper to develop the large spin, large twist systematics at subleading order along the lines of \cite{Alday1} which considered only leading twists. 

\item Our results used the scalar four point function as the starting point. Whether a similar conclusion can be reached by bootstrapping other four point functions of operators with spin $\ell \neq 0$ is an interesting open problem. 

\item Our results agreed exactly with the large-$n$ behaviour found using the Eikonal approximation in AdS/CFT. On the dual gravity side, one can try  to get the subleading terms in $n$ for the case $\ell\gg n$.

\item It will be very interesting to verify our claims for the $n\gg \ell \gg 1$ limit using an effective field theory approach as in \cite{eff}. In that paper it was shown how for the zero spin but large twist form of the anomalous dimension changes due to a massive mode. There it was assumed that there is no stress tensor exchange. To compare with our claims one will need to extend their analysis to arbitrary spin and allowing for a stress tensor exchange.

\item It will be interesting to see if Nachtmann's original proof \cite{Nachtmann:1973mr} can be extended to the $n\neq 0$ case.


\end{itemize}

\section{Acknowledgements}
We thank Fernando Alday, Agnese Bissi, Justin  David, Zohar Komargodski, Tomasz Lukowski, Juan Maldacena, Joao Penedones, Sheer-El-Showk and David Simmons-Duffin for discussions and useful comments. We  thank Agnese Bissi and Tomasz Lukowski for generously sharing their mathematica notebook which helped us understand the results of \cite{Alday}.  We also thank Fernando Rejon-Barrera for pointing out a few typos in our earlier version and confirming several of our results. AS acknowledges support from a Swarnajayanti fellowship, Govt. of India.

\appendix

\section{Calculation details}\label{secA}

To clearly see the expressions for the anomalous dimensions discussed in the main text we now take a mathematical detour a little to explain some of the steps and the useful formulae that goes into the derivation of the above expressions. Note that in the following calculations we will not put the overall factor of $4P_m$ for convenience. Each of the above expressions use the summation of the generic type
\be
a(x,m,\e)= \sum_{k=0}^m \frac{\G(x+k)}{k! \G(x)}\e^k.
\ee
Using the integral representaion of the $\G$-function, the summation on the \textit{rhs} can be converted into,
\be
a(x,m,\e)=\frac{1}{\G(x)}\int_0^\infty dt \ e^{-t}\sum_{k=0}^m \frac{t^{x+k-1}}{k!}\e^k.
\ee
The summation inside the integral can be written as,
\be
\sum_{k=0}^m \frac{t^{x+k-1}}{k!}\e^k=e^{\e t} t^{x-1} \frac{\G(m+1, \e t)}{\G(m+1)}=e^{\e t} t^{x-1}\int_{\e t}^\infty z^m e^{-z}dz\,,
\ee
where $\G(a,x)$ is the incomplete Gamma function given by $\G(a,x)=\int_x^\infty z^{a-1}e^{-z}dz$. Thus the function $a(x,m)$ becomes after the above substitution as,
\be
a(x,m,\e)=\frac{1}{\G(x)\G(m+1)}\int_{0}^\infty dt \ e^{(\e-1)t} t^{x-1}\int_{\e t}^\infty dz \ z^m e^{-z}\,.
\ee 
At this point we do a change of variable from $z$ to $ z=y+\e t$ whereby we notice that the limits of the integral on $z$ changes to $y=0$ and $y=\infty$ respectively. Thus we get,
\be\label{axme}
a(x,m,\e)= \frac{1}{\G(x)\G(m+1)}\int_{0}^\infty\int_0^\infty dt \ dy \ (y+\e t)^m e^{-(t+y)} t^{x-1}\,.
\ee
Whatever summation formulae we have derived in the text are linear combinations of the above function and its derivatives. For example,
\be 
a(x,m,\e=1)=\frac{\G(x)\G(m+x+1)}{\G(x+1)\G(m+1)}\,.
\ee

Again a polynomial arranged like,
\begin{align}\label{poly}
\begin{split}
&\sum_{k=0}^m[c_0+c_1 k+c_2k(k-1)+c_3k (k-1)(k-2)+c_4k(k-1)(k-2)(k-3)+\cdots]\frac{\G(k+x)}{k!\G(x)}\\
&=c_0 a(x,m,\e)|_{\e=1}+c_1\pd_\e a(x,m,\e)|_{\e=1}+c_2\pd_\e^2 a(x,m,\e)|_{\e=1}+c_3\pd_\e^3 a(x,m,\e)|_{\e=1}\\
&+c_4\pd_\e^4 a(x,m,\e)|_{\e=1}+\cdots\,,
\end{split}
\end{align}
where,
\be\label{daxme}
\pd_\e^i a(x,m,\e)|_{\e=1}=\sum_{k=0}^m k(k-1)\cdots(k-i+1) \frac{\G(x+k)}{k!\G(x)}=\frac{\G(m+x+1)}{(x+i)\G(m-i+1)\G(x)}\,.
\ee

\section{Verification of some useful formulae}\label{secB}
With the definitions of the formula in the previous section we can now apply them to our cases specific to the exchange of the twist-$2$ scalar and a spin-$2$, twist-$2$ field.

\subsection{ $\ell_m=0$ and $\tau_m=2$}

We will first deal with the case of a twist-$2$ scalar exchange. The formulae are much simpler for this case.

\begin{enumerate}
\item{ 
\be
\frac{(-m)_k^2(-1-\ell_m+\D-\frac{\ta_m}{2})_k}{(1-\ell_m-m-\frac{\ta_m}{2})_k^2 k!}=\frac{\G(-2+k+\D)}{k!\G(-2+\D)}\,.
\ee
This formula needs no verification. We can simply put $\ell_m=0$ and $\ta_m=2$ to see that the \textit{rhs} is produced. 
}
\item{ 
\be
\sum_{k=0}^m \frac{\G(x+k)}{k!\G(x)}=\frac{\G(1+m+x)}{\G(1+m)\G(1+x)}\,.
\ee
To see this we recall from the previous section that
\be
\sum_{k=0}^m\frac{\G(x+k)}{k!\G(x)}=a(x,m,\e=1)\,.
\ee
Performing the integrals at $\e=1$, fixes the form on the \textit{rhs} of the above formula. 
}
\item{
\be
\g_n=\sum_{m=0}^n a_{n,m}\,,
\ee

In this case the coefficients $a_{n,m}$ are given by,
\be
a_{n,m}=-\frac{(-1)^{m+n}}{8}\frac{(\D_\phi-1)\G(n+1)\G(\D_\phi)\G(2\D_\phi+n+m-3)}{m!(n-m)!\G(\D_\phi+m-1)\G(2\D_\phi+n-3)}\,.
\ee

We will now use the reflection formula for the $\G$-functions to obtain,
\be
\G(m+\D_\phi-1)=(-1)^{-(m+1)}\frac{\pi}{\sin(\pi \D_\phi) \G(2-\D_\phi-m)}\,.
\ee
Separating out the $m$ independent parts and using the integral representation of the product of the $\G$-functions given by,

\be
\G(n+m+2\D_\phi-3)\G(2-\D_\phi-m)=\int_{0}^\infty\int_{0}^\infty dx dy e^{-(x+y)}x^{m+n+2\D_\phi-4}y^{-m+1-\D_\phi}\,,
\ee
we can perform the sum over $m$ to get,
\be
\sum_{m=0}^n (x/y)^m\frac{n!}{m!(n-m)!}=\frac{1}{n!}\bigg(\frac{x+y}{y}\bigg)^n\equiv b(n,x,y)\,.
\ee
Hence the coefficient $\g_n$ associated with the anomalous dimensions become,
\be
\g_n=\frac{(-1)^{n+1}\sin(\pi\D_\phi)}{\pi}\frac{(\D_\phi-1)\G(n+1)\G(\D_\phi)}{8\G(n+2\D_\phi-3)}\int_{0}^\infty dx dy \ b(n,x,y) e^{-(x+y)}x^{n+2\D_\phi-4}y^{1-\D_\phi}\,.
\ee
Using the transformation of variables for $x=r^2 \cos^2\theta$ and $y=r^2\sin^2\theta$ and performing the integral over only the first quadrant, the integration limits change from $r=0$ to $r=\infty$ and $\theta=0$ to $\theta=\pi/2$. The integral thus becomes,
\be
\int_{0}^\infty dx dy \ b(n,x,y) e^{-(x+y)}x^{n+2\D_\phi-4}y^{1-\D_\phi}=-\frac{(-1)^{n-1}\pi\csc(\pi\D_\phi)\G(n+2\D_\phi-3)}{\G(n+1)\G(\D_\phi-1)}\,.
\ee
Putting this with the overall factors we get,
\be
\g_n=-\frac{1}{8}(\D_\phi-1)^2\,.
\ee
which is independent of $n$. Here we have not taken into account the overall factor of $4P_m$ that we should multiply with the expression for $\g_n$ to match the result with the main text.  

}
\end{enumerate}

\subsection{ $\ell_m=2$ and $\tau_m=2$}

We list below the derivation of important formulae required pertaining to this case.

\begin{enumerate}
\item{ 
\be
\frac{(-m)_k^2(-1-\ell_m+\D-\frac{\ta_m}{2})_k}{(1-\ell_m-m-\frac{\ta_m}{2})_k^2 k!}=\frac{(1-k+m)^2(2-k+m)^2\G(-4+k+\D)}{(1+m)^2(2+m)^2\G(1+k)\G(-4+\D)}\,.
\ee
As in the scalar case we put $\ta_m=2$ and $\ell_m=2$ for this case to retrieve the \textit{rhs} of the above formula.
}
\item{ 
\be
\sum_{k=0}^m \frac{(1-k+m)^2(2-k+m)^2}{k!(1+m)^2(2+m)^2}\frac{\G(x+k)}{\G(x)}=\frac{4[6m^2+6m(3+x)+(3+x)(4+x)]\G(3+m+x)}{(1+m)(2+m)\G(3+m)\G(5+x)}\,.
\ee
To get to this, we will appeal to \eqref{poly}, by noticing that the factor $(1-k+m)^2(2-k+m)^2$ can be arranged as,
\beq
(1-k+m)^2(2-k+m)^2=A k (k - 1) (k - 2) (k - 3) + B k (k - 1) (k - 2)\nonumber\\ + C k (k - 1) + 
 D k + E\,,
\eeq
where $A=1$, $B=-4m$, $C=6 m^2 + 6 m + 2$, $D=-4(m+1)^3$ and $E=(2 + 3 m + m^2)^2$. Thus the sum becomes,
\begin{align}
\begin{split}
\sum_{k=0}^\infty \frac{(1-k+m)^2(2-k+m)^2}{(m+1)^2(m+2)^2}\frac{\G(x+k)}{k!\G(x)}=&A \pd_\e^4 a(x,m,\e)|_{\e=1}+B\pd_\e^3 a(x,m,\e)|_{\e=1}\\
&+C\pd_\e^2 a(x,m,\e)|_{\e=1}+D \pd_\e a(x,m,\e)|_{\e=1}\\&+E a(x,m,\e)|_{\e=1}\,.
\end{split}
\end{align}
We know how the each of the terms go by looking at \eqref{daxme}. By combining the coefficients we find that the \textit{rhs} is produced. 
}
\item{
\be
\g_n=\sum_{m=0}^n a_{n,m}\,.
\ee

We will now prove the final piece of the analytic puzzle as follows. First note that $a_{n,m}$ for $\ell_m=2$ and $\ta_m=2$ is given in a closed form expression as
\beq
a_{n,m}=(-1)^{n+m}\frac{15(6m^2+6m(\D_\phi-1)+\D_\phi(\D_\phi-1))}{4\D_\phi}\nonumber\\
\times\frac{\G(n+1)\G(\D_\phi)\G(n+m+2\D_\phi-3)}{m!(n-m)!\G(m+\D_\phi-1)\G(n+2\D_\phi-3)}\,.
\eeq
We will now use the reflection formula for the $\G$-functions to obtain,
\be
\G(m+\D_\phi-1)=(-1)^{-(m+1)}\frac{\pi}{\sin(\pi \D_\phi) \G(2-\D_\phi-m)}\,.
\ee
Separating out the $m$-independent parts we have 
\beq
\g_n=\frac{(-1)^{n+1}\sin(\pi\D_\phi)}{\pi}\frac{15\G(n+1)\G(\D_\phi)}{\G(n+2\D_\phi-3)4\D_\phi}\sum_{m=0}^n \frac{1}{m!(n-m)!}[6m^2+6m(\D_\phi-1)\nonumber\\+\D_\phi(\D_\phi-1)]
\G(n+m+2\D_\phi-3)\G(2-\D_\phi-m)\,.
\eeq
The integral representation of the product of the two $\G$-functions is given by
\be
\G(n+m+2\D_\phi-3)\G(2-\D_\phi-m)=\int_{0}^\infty\int_{0}^\infty dx dy e^{-(x+y)}x^{m+n+2\D_\phi-4}y^{-m+1-\D_\phi}\,.
\ee
Performing the sum over $m$ inside the integral for a polynomial multiplying the $\G$-functions of the form $f(m)=c_0+c_1 m+c_2 m^2$ we get,
\be
\sum_{m=0}^n \bigg(\frac{x}{y}\bigg)^m \frac{f(m)}{m!(n-m)!}=\bigg(\frac{x+y}{y}\bigg)^n\frac{c_0(x+y)^2+c_1 nx(x+y)+c_2nx(nx+y)}{(x+y)^2 n!}\equiv b(n,x,y)\,.
\ee
Thus the expression for $\g_n$ becomes,
\be\label{gn}
\g_n=\frac{(-1)^{n+1}\sin(\pi\D_\phi)}{\pi}\frac{15\G(n+1)\G(\D_\phi)}{\G(n+2\D_\phi-3)4\D_\phi}\int_{0}^\infty dx dy \ b(n,x,y) e^{-(x+y)}x^{n+2\D_\phi-4}y^{1-\D_\phi}\,.
\ee
Using the transformation of variables for $x=r^2 \cos^2\theta$ and $y=r^2\sin^2\theta$ and performing the integral over only the first quadrant, the integration limits change from $r=0$ to $r=\infty$ and $\theta=0$ to $\theta=\pi/2$. Thus, putting the values of $c_0=\D_\phi(\D_\phi-1)$, $c_1=6(\D_\phi-1)$ and $c_2=6$, we have
\begin{align}
\begin{split}
&\int_{0}^\infty dx dy \ b(n,x,y) e^{-(x+y)}x^{n+2\D_\phi-4}y^{1-\D_\phi}\\
&= -\frac{(-1)^{n-1}\pi\csc(\pi\D_\phi)\G(n+2\D_\phi-3)}{\G(n+1)\G(\D_\phi+1)}[6n(n+2\D_\phi-3)(2-\D_\phi+n(n+2\D_\phi-3))\\
&+\D_\phi(\D_\phi-1)(\D_\phi(\D_\phi-1)+6n(n+2\D_\phi-3))]\,.
\end{split}
\end{align}
Multiplying this by the overall $n$-dependent factors outside we have,
\begin{align}
\begin{split}
\g_n=&-\frac{15}{4\D_\phi^2}[6n^4+\D_\phi^2(\D_\phi-1)^2+12n^3(2\D_\phi-3)+6n^2(11-14\D_\phi+5\D_\phi^2)\\
&+6n(2\D_\phi-3)(\D_\phi^2-2\D_\phi+2)]\,,
\end{split}
\end{align}
which is the precise formula for $\g_n$ in $d=4$ dimensions. Note that the final expression for $\g_n$ derived above needs to be multiplied by an overall factor of $4P_m$ to match with that in the main text.
}
\end{enumerate}

\section{$n$ dependence of $\g_n$ for $\ell_m>2$}\label{secC}

In this section we will give an overview on the leading $n$ dependence of the coefficients of the anomalous dimensions \textit{viz.} $\g_n$. We will consider two cases with twist-$2$ and spins $\ell_m=4,6$. For $\ell_m=4$, the coefficients $a_{n,m}$ are given by,
\begin{align}
\begin{split}
a_{n,m}=&-\frac{315P_m(-1)^{m+n}\G(n+1)\G(\D_\phi)^2\G(2\D_\phi+m+n-3)}{\G(m+1)\G(n-m+1)\G(\D_\phi+3)\G(\D_\phi+m-1)\G(2\D_\phi+n-3)}\\
&\times [70m^4+140m^3(\D_\phi-1)+10m^2(9\D_\phi^2-15\D_\phi+11)+10m(2\D_\phi^3-3\D_\phi^2+5\D_\phi-4)\\
&\D_\phi(\D_\phi^2-1)(\D_\phi+2)]\,.
\end{split}
\end{align}
To calculate the leading $n$ dependence in the coefficient $\g_n$, we take the leading term proportional to $m^4$ in $a_{n,m}$ and do the sum over $m$ to get,
\be
\g_n=\sum_{m=0}^n a_{n,m}=-\frac{22050 P_m n^8}{\D_\phi^2(\D_\phi+1)^2(\D_\phi+2)^2}- \cdots\,.
\ee
Thus the leading $n$ dependence of the coefficients $\g_n$ for $\ell_m=4$ is $\sim -n^8$. Similarly for $\ell_m=6$, the coefficients $a_{n,m}$ are given by,
\begin{align}
\begin{split}
a_{n,m}==&-\frac{6006P_m(-1)^{m+n}\G(n+1)\G(\D_\phi)^2\G(2\D_\phi+m+n-3)}{\G(m+1)\G(n-m+1)\G(\D_\phi+5)\G(\D_\phi+m-1)\G(2\D_\phi+n-3)}\\
&\times [924 m^6+2772 m^5 (\D_\phi-1)+210 m^4(15\D_\phi^2-27\D_\phi+26)+420 m^3(\D_\phi-1)\\
&(4\D_\phi^2-5\D_\phi+15)+42m^2(10\D_\phi^4-20\D_\phi^3+95\D_\phi^2-145\D_\phi+88)+42m(\D_\phi^5\\
&15\D_\phi^3-30\D_\phi^2+38\D_\phi-24)+(\D_\phi+4)(\D_\phi+3)(\D_\phi+2)(\D_\phi+1)\D_\phi(\D_\phi-1)]\,.
\end{split}
\end{align}
Again, we take the leading term in $m$ in $a_{n,m}$ and sum over $m$ to get,
\be
\g_n=\sum_{m=0}^n a_{n,m}=-\frac{5549544P_m n^{12}}{\D_\phi^2(\D_\phi+1)^2(\D_\phi+2)^2(\D_\phi+3)^2(\D_\phi+4)^2}-\cdots\,.
\ee
All the above expressions for $\g_n$ are upto overall normalization constants. Thus for a generic $\ell_m$ we find that the coefficient $\g_n$ has an $n$ dependence given by,
\be
\g_n\sim -n^{2\ell_m}\,.
\ee

\section{ Subleading correction at large $\ell$ and large $n$}\label{secD}
In this section we will provide an argument why it is sufficient to consider the expansion of the Bessel functions in \eqref{subu} upto the order we did. To see that, consider the differential equation for the hypergeometric function ${}_2F_1(\b/2,\b/2;\b;1-z)$,
\be
z(1-z)\frac{d^2w}{dz^2}+[1-(\b+1)z]\frac{dw}{dz}-\frac{\b^2}{4}w=0\,.
\ee
Here $\b=\ta+2\ell$. The large $\ell$ limit,is same as the large $\b$ limit. We can then expand the solution in the form,
\be
w=w_0+\frac{1}{\b}w_1+O\bigg(\frac{1}{\b^2}\bigg)\,.
\ee
Consider the change of variables as $y=\b^2z$ in which the differential equation takes the form,
\be
y\bigg(1-\frac{y}{\b^2}\bigg)\frac{d^2 w}{dy^2}+\bigg[1-(\b+1)\frac{y}{\b^2}\bigg]\frac{dw}{dy}-\frac{1}{4}w=0\,.
\ee
The differential equations for the functions $w_0$ and $w_1$ are given by,
\begin{align}
y w''_0+w'_0-w_0=0\,,\nonumber\\
yw''_1+w'_1-w_1-2 y w'_0=0\,.
\end{align}
The solutions are given by,
\be
w_0=c_0 K_0(2\sqrt{y})\,, \ \ \ w_1= f_1(2\sqrt{y})\,,
\ee
where$\sqrt{y}=\b\sqrt{z}$. Thus for large $\ell$ we can expand the full solution $w(y)$ as,
\be
w(y)=c_0K_0(2\sqrt{y})+\bigg(\frac{1}{2\ell}-\frac{n}{2\ell^2}\bigg)f_1(2\sqrt{y})+O\bigg(\frac{1}{\ell^3}\bigg)\,.
\ee
Further now if we consider the expansion of the variable $y$,
\be
w(z)=c_0K_0[(2\ell+\ta)\sqrt{z}]+\bigg(\frac{1}{2\ell}-\frac{n}{2\ell^2}\bigg)f_1[(2\ell+\ta)\sqrt{z}]+O\bigg(\frac{1}{\ell^3}\bigg)\,.
\ee
In the limit $\ell\gg n\gg1$, we have,
\be
w(z)=c_0(K_0(2\ell\sqrt{z})-\sqrt{z}\ta K_1(2\ell\sqrt{z}))+O(n/\ell^2)\,.
\ee
 The terms coming from $w_1$ are hence subleading compared to the leading order result in the limit of large $\ell$. The overall constant $c_0$ is given by,
\be
c_0=\frac{\G(2\ell+\ta)}{\G(\ell+\frac{\ta}{2})^2}=\frac{2^{2\ell+\ta-1}}{\sqrt{\pi}}\ell^{1/2}\bigg(1+\frac{2\ta-1}{8\ell}\bigg)\,.
\ee
Combined with the leading order expansion for $w(z)$ gives \eqref{subu}.

\section{Correction to OPE coefficients for $\ell\gg n\gg1$}\label{sec5}

We now turn to the question about what happens to the leading corrections to the OPE coefficients for the $\ell\gg n\gg1$ case. The starting point of the calculation is,
\be
\sum_{n,\ell}P^{MFT}_{2\D_\phi+2n,\ell}\bigg(\d P_{2\D_\phi+2n,\ell}+\frac{1}{2}\g(n,\ell)\frac{\pd}{\pd n}\bigg)v^n 4^\ell \ell^{1/2}K_0(2\ell\sqrt{z})F^{(4)}[2\D_\phi+2n,v]=\sum_\a A_\a v^\a\,,
\ee
where we are now only considering the terms without the $\log v$ term in \eqref{largel1}. As before we can perform the integration over the spins to eliminate one of the sums. To get the same leading order in $z$ as explained in \cite{anboot}, the coefficients $\d P_{2\D_\phi+2n,\ell}$ should go like,
\be
\d P_{2\D_\phi+2n,\ell}=\frac{\mc_n}{\ell^{\ta_m}}\,.
\ee
Thus the above equation becomes, after performing the $\ell$ integration,
\be
\frac{1}{8}\G\bigg(\D_\phi-\frac{\ta_m}{2}\bigg)^2\sum_n q_{\D_\phi,n}\bigg[\mc_n+\frac{1}{2}\g_n\frac{\pd}{\pd n}\bigg]v^n F^{(4)}[2\D_\phi+2n,v]=\sum_\a A_\a v^\a\,.
\ee
Acting the derivatives of $n$ on $v^n$ obtains a $ v^n\log v$ term and the terms containing only $v^n$ come from considering,
\be\label{eqcn}
\frac{1}{8}\G\bigg(\D_\phi-\frac{\ta_m}{2}\bigg)^2\sum_n q_{\D_\phi,n}\bigg(\mc_n F^{(4)}[2\D_\phi+2n,v]+\frac{1}{2}\g_n\pd_nF^{(4)}[2\D_\phi+2n,v]\bigg)v^n=\sum_\a A_\a v^\a\,.
\ee
At this point note that the function $F^{(4)}[2\D_\phi+2n,v]=2^\ta  {} _2F_1(\D_\phi+n-1,\D_\phi+n-1,2\D_\phi+2n-2;v)$ has a separate $n$ dependent part coming from the $2^\ta$. So the $n$-derivative should act on this part as well.  Thus equation \eqref{eqcn} becomes,
\be\label{ccn}
\frac{1}{8}\G\bigg(\D_\phi-\frac{\ta_m}{2}\bigg)^2\sum_{n=0}^\infty\sum_{k=0}^\infty q_{\D_\phi,n}d_{n,k}(\mc_n+\g_n(\log 2+ g_{n,k}))v^{n+k}=\sum_{\a=0}^\infty A_{\a}v^\a\,.
\ee
where the function $g_{n,k}, d_{n,k}$ are defined as,
\begin{eqnarray}
g_{n,k}&=&\psi(2\D_\phi+2n-2)+\psi(n+\D_\phi+k-1)-\psi(\D_\phi+n-1)-\psi(2\D_\phi+2n+k-2)\,,\\
d_{n,k} &=& \frac{(\D_\phi+n-1)_k^2}{(2\D_\phi+2n-2)_k k!}\,,
\end{eqnarray}
and $\psi(z)=\G'(z)/\G(z)$ is the digamma function. To regroup the terms in \eqref{ccn} increasing powers of $v^\a$, we set $n+k=\a$ and the \textit{lhs} of the above equation becomes $\sum_{\a=0}^\infty f_{\a,\D_\phi}v^\a$ where,
\be\label{eqcn1}
f_{\a,\D_\phi}=\sum_{k=0}^\a q_{\a-k,\D_\phi}d_{\a-k,k}\mc_{\a-k}+b_\a, \ \  \text{where} \ \ b_\a=\sum_{k=0}^\a q_{\a-k,\D_\phi}d_{\a-k,k}\g_{\a-k}(\log 2+g_{\a-k,k}))\,.
\ee
By equating the two sides of the above equation via $f_{\a,\D_\phi}=A_\a$, we can get the coefficients $\mc_n$ once we know the anomalous dimensions $\g_n$. On the \textit{lhs} of \eqref{eqcn}, the coefficients $A_\a$ are determined as follows. We have absorbed the term $(1-v)^{\D_\phi-1}$ in to the \textit{lhs} of \eqref{largel} to obtain,
\begin{align}
\begin{split}
&(1-v)^{\ta_m/2+\ell_m+1-\D_\phi}\frac{P_m\G(\ell_m+2\ta_m)}{4\G(\ell_m+\frac{\ta_m}{2})^2}\sum_{n=0}^\infty\bigg(\frac{(\ell_m+\ta_m/2)_n}{n!}\bigg)^2(2(\psi(n+1)-\psi(\ta_m/2+\ell_m+n))v^n\\&=\sum_{\a=0}^\infty A_\a v^\a\,.
\end{split}
\end{align}
The coefficients $A_\a$ can be written (after transposing the overall factor of $1/8$ to the \textit{rhs} of \eqref{ccn} for the two cases of scalar and spin-$2$ operators as,

\[
 A_\a =
  \begin{cases} 
      \hfill 0    \hfill & \text{ $\ell_m=0$ } \\
      \hfill -2P_m\frac{3\G(\ta_m+2\ell_m)}{\G(\ta_m/2+\ell_m)^2\G(\D_\phi-1)^2}\frac{(\D_\phi+2\a-1)\G(\D_\phi+\a-1)}{\G(\a+1)\G(\D_\phi)} \hfill & \text{ $\ell_m=2$} \\
  \end{cases}
\]

\begin{figure}
\centering
  \includegraphics[width=0.8\textwidth]{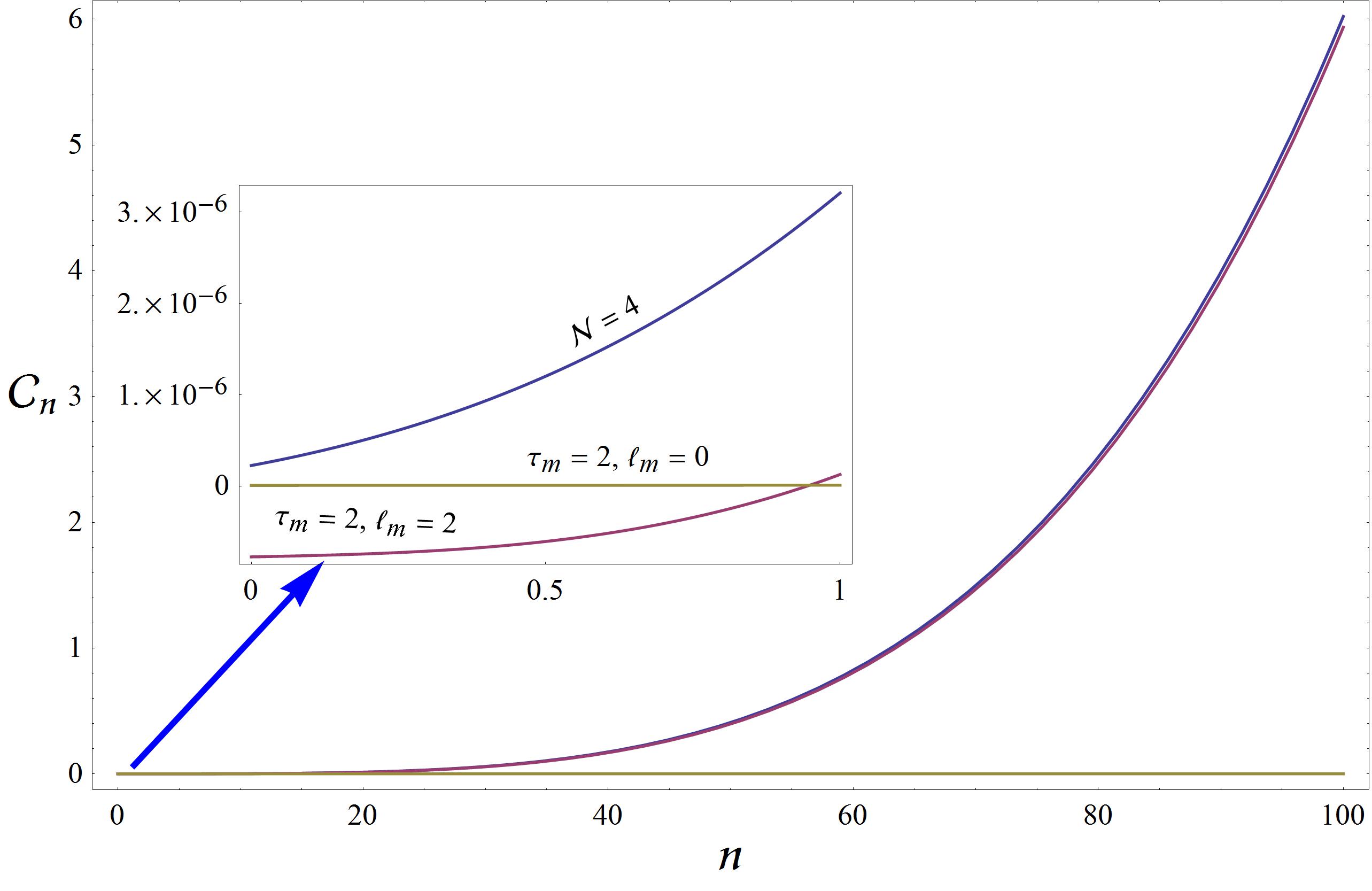}
  \caption{Plot for $\mc_n$ for three cases. The blue curve is for ${\mathcal{N}}=4$, the red curve for the twist-$2$, spin-$2$ operator exchange and the yellow for the twist-$2$ scalar. We have scaled down the OPE coefficients by a factor $10^8$ in this figure.}
  \label{fig:sub1}
\end{figure}

\noindent We can thus write \eqref{eqcn1} as,
\be
\sum_{k=0}^\a q_{\a-k,\D_\phi}d_{\a-k,k}\mc_{\a-k}=A_\a-b_\a\equiv B_\a\,,
\ee
with $b_\a$ given in \eqref{eqcn1}.
This relation can be inverted in the same spirit as we did for the anomalous dimensions. After inversion the corrections to the OPE coefficients can be written as,
\be\label{mcn}
\mc_n=\G(\D_\phi-1)^2\sum_{m=0}^n c_{n,m}B_m\,,
\ee
where we have defined the coefficients $B_\a$ above and $c_{n,m}$ is the same coefficient as given in \eqref{cnm}. Unfortunately to extract a closed form for the coefficients $\mc_n$ from the above sum appears difficult. Nevertheless the behaviour of the OPE corrections can be inferred from \eqref{mcn}. In figure \eqref{fig:sub1} below we have done a comparative study of the OPE corrections for ${\mathcal{N}}=4$ SYM \cite{Alday}, when the \textit{lhs} of \eqref{largel} is dominated by a twist-$2$, spin-$2$ operator and for twist-$2$ scalar operators. From the figure we see that at large $n$, $\mc_n$ tend to follow the relation, 
\be
\mc_n=\frac{1}{2\tq_{\D_\phi,n}}\pd_n(\tq_{\D_\phi,n}\g_n)\,.
\ee 

\noindent whereas for small $n$ there are deviations from the ${\mathcal{N}}=4$ case. From the inset in figure \eqref{fig:sub1} we see that for low lying values of $n$, $\mc_n$ for the twist-$2$, spin-$2$ operator exchange becomes negative while those for the ${\mathcal{N}}=4$ case are positive. $\mc_n$ for the scalar exchange case is a constant positive value. 

We were unable to extend our calculations to the $n\gg \ell \gg 1$ case. The reason is that in order to compute the coefficient $\mc_n$ using the methods in this section we would need to know all the coefficients $\mc_0 \cdots \mc_{n-1}$.  This is not possible since we only know the leading order form of $\gamma_n$ in this limit.

\section{The $n\gg \ell\gg 1$ case}\label{sec4}
We will now turn to the $n\gg \ell\gg 1$ case. This is an interesting limit since the impact parameter in the dual gravity side now is small and one could expect to see non-universality corresponding to contributions from higher spin, higher twist exchanges on the {\it lhs} of the bootstrap equation which correspond to ``stringy" modes. It will not be possible to give a rigorous derivation for the behaviour of the anomalous dimensions in this limit. We will make an ansatz for the anomalous dimension and then using a saddle point approximation extract the behaviour in this limit. It will turn out that this ansatz correctly captures the $\ell \gg n \gg 1$ case and is in exact agreement with the Eikonal calculation in AdS/CFT. First we will make a change of variables and check that these operators exist.

\subsection{Existence of double trace operators for $n\gg\ell\gg1$ limit }

Let us demonstrate that for $n\gg\ell\gg1$, double trace operators exist in large $N$ theories. We first make a change of variables $\bar{h}=\D_\phi+n$, $ h=\bar{h}+\ell$. These are the same variables used in the AdS/CFT Eikonal approximation \cite{Cornalba3}. The reason for making a change of variables to $h,\bar h$ should be obvious--whenever, $n,\ell$ are large, irrespective of which is bigger, $h,\bar h$ are both large and $h\gg \bar h$. The $P^{MFT}$ in terms of these variables is,
\be
P_{MFT}=\frac{2^{7-2(h+\bh)}\pi (h+\bh-2)(h-\bh+1)}{\G(\D_\phi-1)^2\G(\D_\phi)^2}\frac{\G(h)\G(\bh-1)\G(h+\D_\phi-2)\G(\bh+\D_\phi-3)}{\G(h-\frac{1}{2})\G(\bh-\frac{3}{2})\G(h+2-\D_\phi)\G(\bh+1-\D_\phi)}\,.
\ee
For large $h$, the conformal block in the crossed channel is,
\be
g_{h,\bh}(v,u)=\frac{2^{2h-1}}{\sqrt{\pi}}h^{1/2}K_{0}(2h\sqrt{z})v^{\bh}F(\bh,v)\,,
\ee
where,
\be
F(\bh,v)=\frac{1}{1-v}\ {}_2F_1(\bh-1,\bh-1,2\bh-2,v)\,.
\ee
$P_{MFT}$ in the limit of large $h$ can be written as,
\be
P_{MFT}\approx \frac{2^{7-2(h+\bh)}\pi \G(\bh-1)\G(\bh+\D_\phi-3)}{\G(\bh-\frac{3}{2})\G(\bh+1-\D_\phi)\G(\D_\phi-1)^2\G(\D_\phi)^2}h^{2\D_\phi-3/2}\,.
\ee
Combining this together we find,
\begin{align}
\bigg(\frac{u}{v}\bigg)^{\D_\phi}\sum_{h,\bh}P_{MFT}g_{h,\bh}(v,u)&=\frac{4^3\sqrt{\pi}z^{\D_\phi}}{\G(\D_\phi)^2\G(\D_\phi-1)^2}\sum_{\text{large}\ h}h^{2\D_\phi-1}K_0(2h\sqrt{z})\nonumber\\
&\times\sum_{\bh}\frac{4^{-\bh}(1-v)^{\D_\phi}v^{\bh-\D_\phi}\G(\bh-1)\G(\bh+\D_\phi-3)}{\G(\bh-3/2)\G(\bh-\D_\phi+1)}F(\bh,v)\,.
\end{align}
Performing the sum (integral) over large $h$ we get the remaining sum in $\bh$ as,
\begin{align}
A(n,v)&=\bigg(\frac{u}{v}\bigg)^{\D_\phi}\sum_{h,\bh}P_{MFT}g_{h,\bh}(v,u)\nonumber\\
&=\sum_{\bh=\D_\phi}^{\D_\phi+n}\frac{4^{2-\bh}\sqrt{\pi}(1-v)^{\D_\phi-1}v^{\bh-\D_\phi}\G(\bh-1)\G(\bh+\D_\phi-3)}{\G(\bh-3/2)\G(\bh+\D_\phi-1)\G(\D_\phi-1)^2}\ {}_2F_1(\bh-1,\bh-1,2\bh-2,v)\,.
\end{align}
We can expand the sum to arbitrarily high orders in $v$ in mathematica and show that the factor of unity is reproduced on the \textit{lhs}. This gives evidence of the existence of these $n\gg\ell\gg1$ operators.

However this is not enough. These operators in the limit $n\gg\ell\gg1$ must also be consistent with the bootstrap equation even at the subleading order. Next we must argue that the anomalous dimension is small since we are assuming perturbation theory to be able to expand $v^\gamma$. We will find that for a stress tensor exchange, the anomalous dimension goes like $n^3/\ell$. So our results will only be valid in the large $N$ limit with a gap, for some $n<n_{max}$. Since there are all powers of $v$ on the {\it lhs}, the question now becomes how to reproduce all the powers. A natural way would be to argue that the operators above the gap (higher spin modes, ``string modes") will somehow alter the $n$-dependence and allow us to consider any value for $n$ as needed from the {\it lhs}. We will initiate a study of this problem.

\subsection{Subleading bootstrap equation in terms of $h$ and $\bh$}\label{sec41}

Since in the $P^{MFT}$ only particular combinations of $h$ and $\bh$ appear, we will assume a general form of the anomalous dimensions in terms of $h$ and $\bh$ to be,
\be
\g(h,\bh)\propto h^\a \bh^\b (h+\bh)^\c (h-\bh)^\d\,,
\ee
where $\a,\b,\c$ and $\d$ are unknown constants{\footnote{This form is an assumption. In $P_{MFT}$, $h,\bar h, h-\bar h, h+\bar h$ appear so we will make an ansatz that in the large $h,\bar h$ limit, we will get the above form. This form is consistent with the Eikonal result of \cite{Cornalba3} as we will find.}. We expect that the form of the anomalous dimension in the limits $\ell\gg n\gg1$ and $n\gg\ell\gg1$ case will pertain to special cases of the above expression. Using the Stirling approximations for the $\G$-functions,
\be
\G(a+b)\approx \frac{\sqrt{2\pi}}{a^{1/2-b}}\bigg(\frac{a}{e}\bigg)^a\,,
\ee
we can write the $P^{MFT}$ as,
\be
P^{MFT}\overset{h,\bh\rightarrow\infty}{\approx}\frac{2^{7-2(h+\bh)}\pi (h-\bh+1)(h+\bh-2)}{\G(\D_\phi)^2\G(\D_\phi-1)^2}(h\bh)^{2\D_\phi-\frac{7}{2}}\,.
\ee
Further for large $h,\bh$ and $z\rightarrow0$ the conformal blocks in the crossed channel take the form,
\be
g_{h,\bh}(v,u)=2^{2h-1}\frac{h^{1/2}}{\sqrt{\pi}}K_0(2h\sqrt{z})\frac{v^{\bh}}{1-v}\ {}_2F_1(\bh-1,\bh-1,2\bh-2,v)\,.
\ee
Using the anomalous dimensions in terms of $h$ and $\bh$, the \textit{rhs} of the bootstrap equation takes the form, 
\begin{align}
\frac{\sqrt{\pi}z^{\D_\phi}}{\G(\D_\phi)^2\G(\D_\phi-1)^2}\int& d\bh \ 2^{5-2\bh}\bh^{2\D_\phi-7+\b}v^{\bh-\D_\phi}(1-v)^{\D_\phi-1}{}_2F_1(\bh-1,\bh-1,2\bh-2,v)\nonumber\\
&\times \int dh \ h^{2\D_\phi-3+\a}(h+\bh)^{\c+1}(h-\bh)^{\d+1}K_0(2h\sqrt{z})\,.
\end{align}
To sort out the unknown exponents $\a,\b,\c$ and $\d$ we will primarily need the $h$ integral which we write out separately for the convenience of the reader.
\be
\int dh \ h^{2\D_\phi-3+\a}(h+\bh)^{\c+1}(h-\bh)^{\d+1}K_0(2h\sqrt{z})\,.
\ee
For large $h$ such that $h\sqrt{z}\gg1$, we can approximate the Bessel function by,
\be
K_0(2h\sqrt{z})\approx \frac{\sqrt{\pi}}{2h^{1/2}}\frac{e^{-2h\sqrt{z}}}{z^{1/4}}\,.
\ee
Plugging this in the $h$ integral, we can write, 
\be
\frac{\sqrt{\pi}}{2}z^{\D_\phi-1/4}\int dh\ h^{2\D_\phi-7/2+\a}(h+\bh)^{\c+1}(h-\bh)^{\d+1}e^{-2h\sqrt{z}}\,.
\ee
This integral can be solved in the two limits by taking the appropriate approximations of the quantity $h+\bh$. For the limit $\ell\gg n\gg1$ case, $h\gg \bh$ and we can write,
\be
h+\bh\approx h\,,
\ee
whereas for $n\gg\ell\gg1$ case,
\be
h+\bh\approx 2\bh\bigg(1+\frac{h-\bh}{2\bh}\bigg)\approx 2\bh\,.
\ee
The subleading part ($\propto \ell/n$) is neglected in this limit. We will now consider the different limits separately.
\subsection{ $\ell\gg n\gg1$}
In this limit $h\gg\bh$ and hence we can write the $h$ integral as (with $\a+\c+\d=m$),
\be
I(h)=\frac{\sqrt{\pi}}{2}z^{\D_\phi-1/4}\int dh\ h^{2\D_\phi-3/2+m}e^{-2h\sqrt{z}}\,.
\ee
We can consider the entire function as $e^{g(h)}$ where,
\be
g(h)=-2\sqrt{z}+(2\D_\phi-3/2+m)\log h\,.
\ee
The saddle is located at,
\be
h_0=\frac{2\D_\phi-3/2+m}{2\sqrt{z}}\,.
\ee
Using saddle point approximation, we find that $I(h)$ takes the form,
\begin{align}
I(h)=\frac{\sqrt{\pi}}{2}z^{\D_\phi-1/4}\sqrt{\frac{2\pi}{-g''(h_0)}}e^{g(h_0)}&=\frac{\pi}{4} z^{-m/2}\sqrt{2}\bigg(\frac{2\D_\phi-3/2+m}{e}\bigg)^{2\D_\phi-3/2+m}\nonumber\\
&\times(2\D_\phi-3/2+m)^{1/2}2^{-(2\D_\phi-3/2+m)}\,.
\end{align}
Matching the power of $z$ on both sides we see that $m=\a+\c+\d=-\ta_m$. The overall coefficient is the same as what we would get if we replace the Bessel function with its exponential form and expand for large $\D_\phi$. Thus the $h$ integral takes the form,
\be
I(h)=c_{\D_\phi}z^{\ta_m/2}\,,
\ee
where combining with the overall factors,
\be\label{cdp}
c_{\D_\phi,\ta_m}=\frac{\sqrt{\pi}}{2}\frac{2^{1/2+\ta_m-2\D_\phi}\G(2\D_\phi-(1/2+\ta_m))}{\G(\D_\phi)^2\G(\D_\phi-1)^2}\,.
\ee
This is the same overall factor for the $\ell\gg n\gg1$ case if we had replaced the function $K_0(2\ell\sqrt{z})$ with its exponential form and expanded in large $\D_\phi$. Thus the $\bh$ integral becomes,
\be
I(\bh)=\frac{1}{4}c_{\D_\phi,\ta_m}z^{\ta_m/2}\int d\bh \ 2^{7-2\bh}\bh^{2\D_\phi-7/2+\b}v^{\bh-\D_\phi}(1-v)^{\D_\phi-1}{}_2F_1(\bh-1,\bh-1,2\bh-2,v)
\ee
We can now convert this into the summation form by noting that the factor $h^{2\D_\phi-7/2}2^{7-2\bh}$ is the asymptotic form of,
\be
q_{\D_\phi,n}=\frac{8\G(\D_\phi+n-1)^2\G(n+2\D_\phi-3)}{\G(n+1)\G(2\D_\phi+2n-3)\G(\D_\phi)^2\G(\D_\phi-1)^2}\overset{n\gg1}{\approx}\frac{n^{2\D_\phi-7/2}2^{7-2\bh}}{\G(\D_\phi)^2\G(\D_\phi-1)^2}\,,
\ee
where for $n\gg\D_\phi$ we can take $\bh=\D_\phi+n\approx n$. We can further replace $\bh^\b$ by $\g_n$ by the $\g_n$ part of $\g(n,\ell)$. Apart from this the other factors in the $\bh$ integral are exactly the same as for the $n$ summation. Finally,
\be
\frac{1}{4}\G(\D_\phi)^2\G(\D_\phi-1)^2c_{\D_\phi,\ta_m}z^{\ta_m/2}\sum_n^\infty \g_n q_{\D_\phi,n}v^n(1-v)^{\D_\phi-1}F^{(d)}(2\D_\phi+2n,v)=lhs\,.
\ee
This summation thus reproduces the correct $n$ dependence of the $\g_n$ functions for this limit as we saw earlier. This argument also fixes the overall sign of the anomalous dimension to be negative.

\subsection{$n\gg\ell\gg1$}
In this limit we will neglect the term $(h-\bh)/2\bh\gg1$. Thus the $h$ integral takes the form,
\be
I(h)=\int dh \ h^{2\D_\phi-2+\a+\d}K_0(2h\sqrt{z})\,.
\ee
Here we have also approximated $h-\bh$ by $h$ since we are still in the limit of large $\ell$. We will further approximate the Bessel function by its exponential form and consider the entire function as $e^{g(h)}$ where,
\be
g(h)= -2\sqrt{z}+(2\D_\phi-5/2+\a+\d)\log h\,.
\ee
Equating $g'(h)=0$ gives the location of the saddle ($\a+\d=p$),
\be
h_0=\frac{2\D_\phi-5/2+p}{2\sqrt{z}}\,.
\ee
Thus,
\begin{align}
I(h)=\frac{\sqrt{\pi}}{2}z^{\D_\phi-1/4}\sqrt{\frac{2\pi}{-g''(h_0)}}e^{g(h_0)}&=\frac{\pi}{4} z^{-(p-1)/2}\sqrt{2}\bigg(\frac{2\D_\phi-5/2+p}{e}\bigg)^{2\D_\phi-5/2+p}\nonumber\\
&\times(2\D_\phi-5/2+p)^{1/2}2^{-(2\D_\phi-5/2+p)}\,.
\end{align}
Thus here $p=\a+\d=1-\ta_m$ and further, the overall coefficient is the same $c_{\D_\phi,\ta_m}$ defined in \eqref{cdp}. Thus,
\be
I(\bh)=\frac{1}{2}c_{\D_\phi,\ta_m}z^{\ta_m/2}\int d\bh \ 2^{7+\c-2\bh}\bh^{2\D_\phi-5/2+\b+\c}v^{\bh-\D_\phi}(1-v)^{\D_\phi-1}{}_2F_1(\bh-1,\bh-1,2\bh-2,v)
\ee
Again from our previous discussion we can convert this integral into a summation giving the required behaviour for the limit $n\gg\ell\gg1$. 

Note that using the two relations,
\be
\a+\c+\d=-\ta_m\,, \ \ \text{and}\ \ \a+\d=1-\ta_m\,, \ \ \text{gives} \ \ \c=-1\,,
\ee 
for both the limits. Thus we have partially fixed the form of the anomalous dimension to be,
\be
\g(h,\bh)\sim-\frac{(\ell+n)^\a n^\b\ell^\d}{(\ell+2n)}\,.
\ee
>From calculating the subleading terms for $\ell\gg n\gg1$ the results from the previous section gives us,
\be
\g(h,\bh)\sim-\frac{n^\b}{\ell^{1-\a-\d}}\bigg(1+(\a-2)\frac{n}{\ell}\bigg)\,,
\ee
so that $\a=2-\ta_m$ which further gives $\d=-1$. Thus  in terms of $n$ and $\ell$,
\be
\g(n,\ell)\sim-\frac{n^\b (\ell+n)^{2-\ta_m}}{\ell(\ell+2n)}\,.
\ee
The remaining exponent can be obtained (see appendix \ref{secE}) by calculating the leading $n$ dependences for various twists. It works out to be,
\be
\b=-2+2\ell_m+\ta_m\,.
\ee
Thus the full expression for the anomalous dimension in the large $h,\bar h$ limit in terms of $\ell, n$, modulo overall factors, takes the form,
\be
\g(n,\ell)\sim -\frac{n^{-2+2\ell_m+\ta_m}(\ell+n)^{2-\ta_m}}{\ell(\ell+2n)}\,.
\ee
This expression is precisely what emerges from the Eikonal approximation in AdS/CFT \cite{Cornalba1} for a generic spin exchange.
We can see that for the two different limits being considered in our paper, it takes the following forms at the leading order,
\be
\g(n,\ell)\overset{\ell\gg n}{\sim} -\frac{n^{-2+2\ell_m+\ta_m}}{\ell^{\tau_m}}\,,\  \ \g(n,\ell)\overset{n\gg\ell}{\sim} -\frac{n^{2\ell_m-1}}{2\ell}\,.  
\ee
Note that in the $n\gg \ell \gg 1$ case the form just depends on the spin with no dependence on the twist. Moreover, the $\ell$-dependence is independent of both $\tau_m, \ell_m$. 
In particular, in the limit $n\gg\ell\gg1$ if we have the leading contribution as that coming from a stress tensor\footnote{Note that the problem that we allude to in this paragraph does not arise for the $\ell_m=0$ case. This is reminiscent of the discussion in  \cite{Camanho:2014apa} where the polarization of the graviton was crucial for the causality arguments. } for large $N$ we have
\be
\gamma(n,\ell) \sim -\frac{n^3}{\ell}\left(\frac{1}{N^2}+\sum_{\ell_m=2}^\infty n^{2\ell_m-4} P^{(m)}_{\ell_m}[1+O(\frac{\ell}{n})]\right)
\ee
where $P^{(m)}_{\ell_m}$ are related to the square of the OPE coefficients for massive spin-$\ell_m$ ($\ell_m$ even) modes on the {\it lhs} of the bootstrap equation. We are assuming that we have added generic spin and twist on the {\it lhs} so that to produce the appropriate powers of $u,v$, namely $u^{\tau_m/2} v^n \log v$ with $n$ being a natural number\footnote{Of course, for specific values of $\tau_m$ these subleading powers of $u$ will also mix with subleading $u$-powers arising from some leading twist. We are ignoring this possibility.}, we will need to modify the form of the anomalous dimension to what we have indicated above. The leading $1/N^2$ dependence thus will be the leading contribution only if the $P^{(m)}_{\ell_m}$'s suppress the contributions from the positive powers of $n$. In other words for this result to hold there has to be a gap in the spectrum with the contributions from operators above the gap being suppressed. Evidently, this suppression will only work for the $n$-dependent operators below the gap. The $O(\ell/n)$ terms will depend on $\tau_m,\ell_m$ and are small in this limit. As we keep increasing $n\sim O(N)$, the assumption that the anomalous dimensions are small will break down (due to the negative sign, one can also be in danger of violating unitarity but this cannot be concluded yet since the anomalous dimension result cannot be trusted when this happens). The interesting question is if adding a single (or a finite number of) higher spin (massive) operator(s) can make the anomalous dimensions small again. The form we have derived above suggests that this is not possible.
If we insisted that the operators for all $n$ have perturbatively small anomalous dimensions, this can only be possible if we resum the contributions from the higher spin modes. The above result seems to suggest that for this to happen one will need an infinite number of higher spin modes since each contribution from the higher spin modes comes with a positive power of $n$. A more complete analysis of this very important problem is however beyond the scope of this paper (for instance at the level of what we have done we cannot say what the OPE coefficients are for us to be able to resum the series).

\section{general $\tau_m$ and $\ell_m$}\label{secE}
\begin{figure}
\centering
  \includegraphics[width=0.8\textwidth]{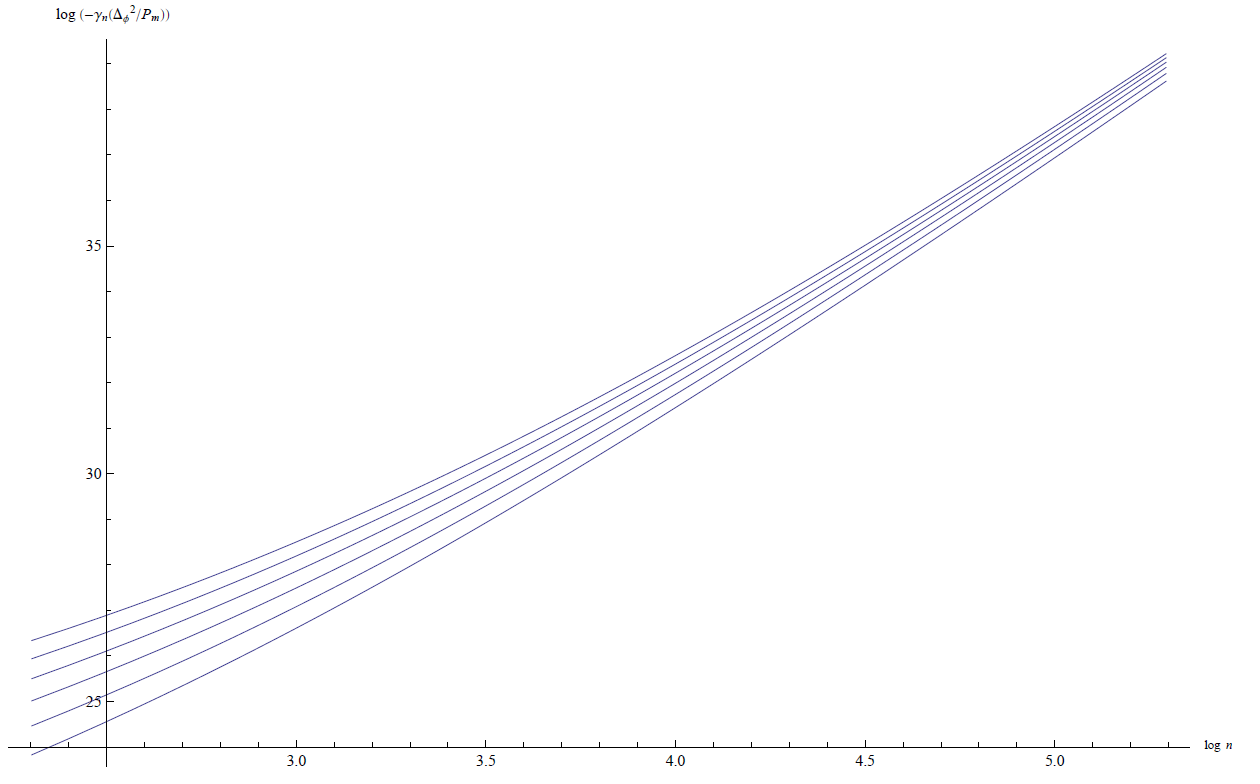}
  \caption{Plot for the numerical estimate of the exponent of $n$ for $\ta_m=4$ for a range of $\D_\phi$. }
  \label{fig:sub2}
\end{figure}

The remaining exponent $\b$ for $n$ in the expression for the anomalous dimension is determined by doing the following exercise. To start with the anomalous dimension looks like,
\be
\g(n,\ell)\sim-\frac{n^\b(\ell+n)^{2-\ta_m}}{\ell(\ell+2n)}\,.
\ee
In the limit $\ell\gg n\gg1$, the anomalous dimensions take the form,
\be
\g(n,\ell)\sim-\frac{n^\b}{\ell^2}\,.
\ee
We can assume a form of $\b=a+b \ell_m+c\ta_m$. Putting this back in and calculating for $\ta_m=2$ and  $\ell_m=2$ (stress tensor) and $\ell_m=0$ (scalar) for which $\b=4$ and $0$ respectively, we get the leading term,
\be
a+2b+2c=4\,, \ a+2c=0\,, \Rightarrow b=2\,.
\ee
To find the other coefficients $a$ and $c$ we need one more data point. In the plot \eqref{fig:sub2} for $\ta_m=4$ and $\ell_m=2$, $\b=6$. Thus,
\be
a+2b+4c=6\,,\Rightarrow c=2\, \ \text{and} \ a=-2\,.
\ee
Thus with $\ell_m=\D_m-\ta_m$ for the minimal twist operator,
\be
\b=-2+2\D_m-\ta_m\,.
\ee


\begin{thebibliography}{99}

\bibitem{Rattazzi}
R.~Rattazzi, V.~S.~Rychkov, E.~Tonni and A.~Vichi,
JHEP {\bf 0812}, 031 (2008)
[arXiv: 0807.0004].\\
R.~Rattazzi, S.~Rychkov, and A.~Vichi,
Phys. \ Rev. \ D {\bf 83} 046011 (2011)
[arXiv: 1009.2725].\\
R.~Rattazzi, S.~Rychkov, and A.~Vichi,
J. \ Phys. \ A {\bf 44} 035402 (2011)
[arXiv: 1009.5985].\\
D.~Pappadopulo, S.~Rychkov, J.~Espin and R.~Rattazzi,
Phys. \ Rev. \ D {\bf 86} 105043 (2012)
[arXiv: 1208.6449].\\
  D.~Poland, D.~Simmons-Duffin and A.~Vichi,
  JHEP {\bf 1205}, 110 (2012)
  [arXiv:1109.5176 [hep-th]].\\
  Y.~Nakayama and T.~Ohtsuki,
  Phys.\ Lett.\ B {\bf 734}, 193 (2014)
  [arXiv:1404.5201 [hep-th]].

\bibitem{Polchinski}
I.~Heemskerk, J.~Penedones, J.~Polchinski and J.~Sully,
JHEP {\bf 0910} (2010) 079
[arXiv: 0907.0151[hep-th]].

\bibitem{Hogervorst}
M.~Hogervorst, H.~Osborn and S.~Rychkov,
JHEP {\bf 1308} (2013) 014
[arXiv: 1305.1321[hep-th]].\\
M.~Hogervorst and S.~Rychkov,
Phys.\ Rev. \ D {\bf D87} (2013) 106004
[arXiv: 1303.1111[hep-th]].

\bibitem{Dolan}
F.~A.~Dolan and H.~Osborn,
Nucl. \ Phys. \ B {\bf 629} (2002) 3-73 
[hep-th/0011040].\\
F.~A.~Dolan and H.~Osborn,
 Annals \ Phys.  {\bf 321} (2006) 581-626
[hep-th/0309180].\\
F.~A.~Dolan and H.~Osborn,
[arXiv: 1108.6194[hep-th]].

\bibitem{SCFTs}
C.~Beem, L.~Rastelli and F.~Passerini,
Phys.\ Rev.\ Lett.\ {\bf 111} (2013) 071601
[arXiv: 1304.1803[hep-th]].\\
C.~Beem, M,~Lemos, P.~Liendo, L.~Rastelli and B.~C.~ van Rees,
[arXiv: 1412.7541[hep-th]].\\
F.~A.~Dolan, M.~Nirschl and H.~Osborn,
Nucl. \ Phys.\ B\ {\bf 749} (2006) 109-152
[hep-th/0601148].\\
F.~A.~Dolan and H.~Osborn,
Nucl. \ Phys.\ B\ {\bf 629} (2002) 3-73
[hep-th/0112251].\\
F.~A.~Dolan and H.~Osborn,
Nucl. \ Phys.\ B\ {\bf 593} (2001) 599-633
[hep-th/0006098].

\bibitem{Showk}
S.~El-Showk, M.~F.~Paulos, D.~Poland, S.~ Rychkov, D.~Simmons-Duffins and A.~Vichi,
Phys.\ Rev.\ D\ {\bf 86}(2012) 025022
[arXiv: 1203.6064[hep-th]].\\
S.~El-Showk, M.~F.~Paulos, D.~Poland, S.~ Rychkov, D.~Simmons-Duffins and A.~Vichi,
J.\ Stat.\ Phys.\ {\bf 157}(2014) 869
[arXiv: 1403.4545[hep-th]].\\
F.~Kos, D.~Poland, D.~Simmons-Duffin,
JHEP \ {\bf 1411} (2014) 109
[arXiv: 1411.7932[hep-th]].\\
F.~Gliozzi and A.~Rago
JHEP \ {\bf 1410} (2014) 42
[arXiv: 1403.6003[hep-th]].

\bibitem{anboot} 
  A.~L.~Fitzpatrick, J.~Kaplan, D.~Poland and D.~Simmons-Duffin,
  JHEP {\bf 1312}, 004 (2013)
  [arXiv:1212.3616 [hep-th]].
\bibitem{Komargodski} 
  Z.~Komargodski and A.~Zhiboedov,
  JHEP {\bf 1311}, 140 (2013)
  [arXiv:1212.4103 [hep-th]].
 \bibitem{Alday2} L.~F.~Alday and J.~M.~Maldacena,
  JHEP {\bf 0711}, 019 (2007)
  [arXiv:0708.0672 [hep-th]].

\bibitem{Nachtmann:1973mr} 
  O.~Nachtmann,
  Nucl.\ Phys.\ B {\bf 63}, 237 (1973).

\bibitem{Vos} 
  G.~Vos,
  arXiv:1411.7941 [hep-th].

\bibitem{Cornalba1} 
  L.~Cornalba, M.~S.~Costa and J.~Penedones,
  JHEP {\bf 0709}, 037 (2007)
  [arXiv:0707.0120 [hep-th]].

\bibitem{Cornalba2} 
  L.~Cornalba, M.~S.~Costa, J.~Penedones and R.~Schiappa,
  Nucl.\ Phys.\ B {\bf 767}, 327 (2007)
  [hep-th/0611123].

\bibitem{Cornalba3} 
  L.~Cornalba, M.~S.~Costa, J.~Penedones and R.~Schiappa,
  JHEP {\bf 0708}, 019 (2007)
  [hep-th/0611122].

\bibitem{Camanho:2014apa} 
  X.~O.~Camanho, J.~D.~Edelstein, J.~Maldacena and A.~Zhiboedov,
  arXiv:1407.5597 [hep-th].

\bibitem{Alday} 
  L.~F.~Alday, A.~Bissi and T.~Lukowski,
  arXiv:1410.4717 [hep-th].

\bibitem{Arutyunov:2002fh} 
  G.~Arutyunov, F.~A.~Dolan, H.~Osborn and E.~Sokatchev,
  Nucl.\ Phys.\ B {\bf 665}, 273 (2003)
  [hep-th/0212116].

\bibitem{eff} 
  A.~L.~Fitzpatrick, E.~Katz, D.~Poland and D.~Simmons-Duffin,
  JHEP {\bf 1107}, 023 (2011)
  [arXiv:1007.2412 [hep-th]].


\bibitem{Bissi}
L.~F.~Alday and A.~Bissi,
JHEP {\bf 1310} (2013) 202
[arXiv: 1305.4604[hep-th]].

\bibitem{Alday1}
L.~F.~Alday, A.~Bissi and T.~Lukowski,
[arXiv: 1502.07707[hep-th]].


\bibitem{boot2} 
  A.~Kaviraj, K.~Sen and A.~Sinha,
  arXiv:1504.00772 [hep-th], to appear in JHEP.


\end{thebibliography}
\end{document}